\documentclass[aps,prd,showpacs,eqsecnum,twocolumn,superscriptaddress,a4paper,nofootinbib]
{revtex4-2}
\usepackage{amsmath,amssymb,graphicx,color}
\usepackage{bm}
\usepackage{epsf}
\usepackage{mathrsfs}

\begin{document}

\title{Long-term evolution of neutron-star merger remnants in general relativistic resistive-magnetohydrodynamics with a mean-field dynamo term}

\author{Masaru Shibata} 
\affiliation{Max Planck Institute for
  Gravitational Physics (Albert Einstein Institute), Am M{\"u}hlenberg 1,
  Potsdam-Golm 14476, Germany}
\affiliation{Center for Gravitational Physics, Yukawa Institute for Theoretical
  Physics, Kyoto University, Kyoto, 606-8502, Japan}

\author{Sho Fujibayashi} 
\affiliation{Max Planck Institute for
  Gravitational Physics (Albert Einstein Institute), Am M{\"u}hlenberg 1,
  Potsdam-Golm 14476, Germany}


\author{Yuichiro Sekiguchi} \affiliation{Department of Physics, Toho
  University, Funabashi, Chiba 274-8510, Japan}
\affiliation{Center for Gravitational Physics, Yukawa Institute for Theoretical
  Physics, Kyoto University, Kyoto, 606-8502, Japan}

\date{\today}
\newcommand{\beq}{\begin{equation}}
\newcommand{\eeq}{\end{equation}}
\newcommand{\beqn}{\begin{eqnarray}}
\newcommand{\eeqn}{\end{eqnarray}}
\newcommand{\pa}{\partial}
\newcommand{\vp}{\varphi}
\newcommand{\varep}{\varepsilon}
\newcommand{\ep}{\epsilon}
\newcommand{\comp}{(M/R)_\infty}
\def\cL{\mathscr{L}}
\newcommand{\cE}{{\cal{E}}}
\newcommand{\cB}{{\cal{B}}}
\newcommand{\cb}{{\cal{H}}}
\newcommand{\cJ}{{\cal{J}}}
\newcommand{\cC}{{\cal{C}}}
\newcommand{\cI}{{\cal{I}}}
\begin{abstract}

Long-term neutrino-radiation resistive-magnetohydrodynamics
simulations in full general relativity are performed for a system
composed of a massive neutron star and a torus formed as a remnant of
binary neutron star mergers.  The simulation is performed in axial
symmetry incorporating a mean-field dynamo term for a hypothetical
amplification of the magnetic-field strength.  We first calibrate the
mean-field dynamo parameters by comparing the results for the
evolution of black hole-disk systems with viscous hydrodynamics
results.  We then perform simulations for the system of a remnant
massive neutron star and a torus. As in the viscous hydrodynamics
case, the mass ejection occurs primarily from the torus surrounding
the massive neutron star. The total ejecta mass and electron fraction
in the new simulation are similar to those in the viscous
hydrodynamics case. However, the velocity of the ejecta can be
significantly enhanced by magnetohydrodynamics effects caused by
global magnetic fields.


\end{abstract}
\pacs{04.25.D-, 04.30.-w, 04.40.Dg}
\maketitle


\section{Introduction}\label{sec1}

The first observation of the binary neutron star merger
GW170817~\cite{GW170817,GW170817a} showed that theoretical modeling
for the merger and post-merger phases of binary neutron stars is the
key for extracting valuable information from the observed
electromagnetic signals. Although there are a variety of the
possibilities for the remnants of the binary neutron star
mergers~\cite{SH19} and for the corresponding electromagnetic
counterparts~\cite{MB2012,HP2015}, irrespective of the possibilities,
the key phenomenon for the strong electromagnetic emission is the mass
ejection from the remnant including the ultra-relativistic jets. Thus,
the important aspect of the theoretical study is to clarify how the
matter is ejected from the system and to understand the properties of
the ejected matter such as total mass, typical velocity, and typical
elements. This motivation has stimulated many numerical simulations
for the merger phase (e.g.,
Refs.~\cite{Rosswog,Rosswog1,PNR2012,Hotoke,Sekig,Foucart16,Radice16,Lehner2016,Bovard,Foucart,Shibata17,Kyutoku})
and for the post-merger phase~(e.g.,
Refs.~\cite{MF2013,MF2014,Perego,Just2015,SM17,FTQFK18,Fujiba17,Fujiba18,Janiuk19,FTQFK19,Miller19,Fujiba20,Fujiba2020,Fujiba20b,FFL20,moesta,SFS2021,Just2021})
in the last decades.


For the binary neutron star merger resulting in the formation of a
massive neutron star, the major mass ejection is likely to occur in
the post-merger phase~\cite{Shibata17,Perego17} and the key process
for the mass ejection from the merger remnant is the
magnetohydrodynamical effect.  For disks (or tori) surrounding the
central compact object (either a massive neutron star or a black
hole), which has approximately Keplerian rotational profile, it is
believed that the magnetorotational instability (MRI)~\cite{BH98} is
activated, and as a result, a turbulence is developed, enhancing the
turbulence viscosity.  The resulting viscous heating and angular
momentum transport inevitably enhance the activity of the disk, and
eventually, the mass ejection is induced.  In addition, the amplified
magnetic field could eventually develop a global magnetic field,
which could further enhance the mass ejection
efficiency~\cite{SM17,FTQFK19,Miller19,Just2021}.  However, to fully
clarify these processes, we need high-resolution simulations in three
spatial dimension that can resolve the unstable modes of the MRI with
a sufficient accuracy. However, to date, due to the limitation of the
computational resource, such expensive simulations have not been
performed yet (but see Ref.~\cite{LTQ} for the high-resolution
simulation to model accretion disks around a supermassive black hole).


The remnant massive neutron star also could be the source for the mass
ejection~\cite{Fujiba18,Fujiba2020,moesta,SFS2021}.  In contrast to
the accretion disk around the compact objects, the angular velocity of
the remnant massive neutron stars, $\Omega(\varpi)$ with $\varpi$
denoting the cylindrical radius, increases with $\varpi$ in the
central region reflecting the nearly irrotational velocity field of
the pre-merger stage of two neutron stars (e.g., Ref.~\cite{STU2005}).
Thus, the major region except for the outermost part of the remnant
neutron star is stable to the MRI.  However, there still exists the
differential rotation in the remnant neutron star, which causes the
winding of the magnetic field and increases the toroidal-field
strength~\cite{Shapiro2000,SLSS2006,Sun2019,SFS2021}. In the presence
of the resultant strong toroidal field, several magnetohydrodynamics
(MHD) instabilities such as the Parker and Taylor
instabilities~\cite{Parker,Taylor} together with the convection and
circulation can take place, and the magnetic-field strength could be
amplified through the dynamo action.  To accurately investigate this
amplification process, we need a high-resolution and long-term MHD
simulation in full general relativity with the relevant microphysics
such as the neutrino transport that can induce the convection and
dynamo~\cite{Guilet,masada}.  However, this is still a formidable work
in the current computational resources.

In this paper, we attack this problem phenomenologically, bypassing
the high-resolution simulation.  We perform general-relativistic
neutrino-radiation resistive-MHD (GRRRMHD) simulations for a massive
neutron star formed as a remnant of equal-mass binary neutron star
mergers in axial symmetry as in our previous paper~\cite{SFS2021}, but
taking into account the mean-field dynamo
term~\cite{B2005,BDZ2013,SFS2021}; i.e., we incorporate the mean-field
dynamo term in the current density $j^\mu$ which is proportional to
the magnetic field.  In the assumption of axial symmetry with no
dynamo term, the poloidal magnetic field is not amplified even when
the toroidal-field strength is significantly enhanced, due to the
anti-dynamo property~\cite{Cowling}. With the phenomenological
incorporation of the dynamo term, by contrast, the poloidal magnetic
field can be amplified in the presence of the toroidal field. As a
result, the magnetic field continues to be amplified even if we start
initially from a purely toroidal magnetic field in the axisymmetric
simulation. In the context of the MHD evolution of the
binary-neutron-star merger remnant, the differential rotation is
likely to be the key for the magnetic-field amplification. Thus, we
pay attention only to the $\alpha$-$\Omega$ dynamo for the
hypothetical field amplification in this paper. 

Similar numerical experiments in MHD have been recently performed by
several groups~\cite{Bugli,Sadowski2015,Tomei,Vourellis} for the
systems of a black hole and a disk assuming the fixed back ground of
the black hole. These works have illustrated that with the
incorporation of the mean-field dynamo term (i.e., with the
$\alpha$-$\Omega$ dynamo effect), the numerical results by the
three-dimensional ideal MHD simulations are at least qualitatively
reproduced. These results encourage us to perform this type of a
phenomenological simulation to capture a realistic MHD evolution
process of the remnant of the binary neutron star mergers, which
cannot be currently studied in the first-principle MHD simulation due
to the poor grid resolution resulting from the limitation of the
computational resources. This type of the phenomenological simulations
is also helpful to explore a wide variety of the possibilities for the
long-term evolution process of the binary-neutron-star merger
remnants, which have not been well explored yet, with a (relatively)
inexpensive computational cost. In particular, to derive theoretical
models for the ejecta with which electromagnetic signals are studied,
the simulation results provide the useful data for the post-process
calculations~(see, e.g.,
Refs.~\cite{Kawaguchi18,Hotoke18,Kawaguchi2021}).

The paper is organized as follows: In Sec.~\ref{sec2}, we summarize
the basic equations employed in the present numerical simulations
paying particular attention to the resistive MHD equations in general
relativity.  In Sec.~\ref{sec3} we first calibrate our method by
performing the simulations for the system of a black hole and a
disk. We compare the results with those in viscous
hydrodynamics~\cite{Fujiba20,Fujiba20b} and confirm that the new
results by the GRRRMHD simulations are quantitatively similar to those
by the previous viscous hydrodynamics simulations with an appropriate
choice of the dynamo parameters, although the MHD effects can modify
the process of the mass ejection.  Then, in Sec.~\ref{sec4}, we
perform a GRRRMHD simulation for a remnant of a binary neutron star
merger composed of a massive neutron star and a torus. By comparing
the results with those by the viscous hydrodynamics
simulation~\cite{Fujiba2020}, we show additionally significant
magnetohydrodynamics effects on the matter ejected from the system.
Section~\ref{sec5} is devoted to a summary.  Throughout this paper, we
use the geometrical units of $c=1=G$ where $c$ and $G$ denote the
speed of light and the gravitational constant, respectively (but $c$
is often recovered to clarify the units in the following
sections). Latin and Greek indices denote the space and spacetime
components, respectively. In Sec.~\ref{sec2}, we suppose to use
Cartesian coordinates for the spatial components whenever equations
are written.

\section{Basic equations for numerical computations}\label{sec2}

\subsection{Brief summary}\label{sec2-1}

We perform a resistive MHD simulation in full general relativity using
the same formulation and numerical implementation as in our previous
paper~\cite{SFS2021} (see this reference for details of the
formulation and numerical methods, and for the results of test-bed
problems).  Specifically, we numerically solve Einstein's equation,
resistive MHD equations incorporating a mean-field dynamo term,
evolution equations for the lepton fractions including the electron
fraction, and (approximate) neutrino-radiation transfer
equations. Except for an MHD part (see the next paragraphs), the basic
equations and input physics are the same as before: Einstein's
equation is solved using the original version of the
Baumgarte-Shapiro-Shibata-Nakamura formalism~\cite{BSSN} together with
the puncture formulation~\cite{puncture}, Z4c constraint propagation
prescription~\cite{Z4c}, and 5th-order Kreiss-Oliger dissipation.  The
axial symmetry for the geometric variables is imposed using the
cartoon method~\cite{cartoon,cartoon2} with the 4th-order accurate
Lagrange interpolation in space.  The lepton fractions are evolved
taking into account electron and positron captures, electron-positron
pair annihilation, nucleon-nucleon bremstrahlung, and plasmon
decay~\cite{Fujiba18,Fujiba2020}.  The same tabulated equation of
state as in Refs.~\cite{Fujiba20,Fujiba20b,SFS2021} is employed.
Specifically, we employ the DD2 equation of state~\cite{DD2} for a
relatively high-density part and the Timmes (Helmholtz) equation of
state for a low-density part~\cite{Timmes}.

We evolve weighted electric and magnetic fields defined,
respectively, by~\cite{SS2005,Shibata16,SFS2021}
\beqn
\cE^\mu &:=& \sqrt{\gamma} F^{\mu\nu} n_\nu,\\
\cB^\mu &:=& 
   {1 \over 2} \sqrt{\gamma} n_\alpha \epsilon^{\alpha\mu\nu\beta} F_{\nu\beta},
\eeqn
where the electromagnetic tensor is written as
\beqn
F^{\mu\nu}=n^{\mu} E^{\nu} - n^{\nu} E^{\mu}
+n_\beta\epsilon^{\beta\mu\nu\alpha} B_{\alpha}, \label{eq:Fab}
\eeqn
with $n^\alpha$ the time-like unit normal vector, $\gamma$ the
determinant of the spatial metric $\gamma_{ij}$, and
$\epsilon^{\mu\nu\alpha\beta}$ the Levi-Civita tensor.  $E^\mu$ and
$B^\mu$ are the electric and magnetic fields in the inertial frame,
respectively.

The evolution equations for $\cE^i$ and $\cB^i$ are written as~\cite{SFS2021}
\beqn
&&\pa_t \cE^i = - \pa_k \left(\beta^i \cE^k - \beta^k \cE^i
+ \alpha \epsilon^{kij} \cB_j \right) \nonumber \\
&& ~~~~~~~~~~~~-4\pi \left(\cJ^i-Q \beta^i \right), \label{eq:emevo1a}\\
&&\pa_t \cB^i = - \pa_k \left( \beta^i \cB^k - \beta^k \cB^i
- \alpha \epsilon^{kij} \cE_j \right) \nonumber \\
&& ~~~~~~~~~~~-\alpha\gamma^{1/2} \gamma^{ij} \pa_j \phi_B\label{eq:emevo2a} \\
&&\pa_t \phi_B=\beta^k \pa_k \phi_B - \alpha \kappa \phi_B-\alpha \gamma^{-1/2} \pa_k \cB^k,
\label{eq:emevo3a}
\eeqn
where $\alpha$ is the lapse function, $\beta^i$ is the shift vector,
$\epsilon^{kij}:=n_\mu \epsilon^{\mu kij}$, $\phi_B$ is a new
auxiliary variable associated with the divergence cleaning, $\kappa$
is a constant, and the current term, $\cJ^i-Q\beta^i$, is written
as~\cite{BDZ2013,Bugli,Vourellis,SFS2021}
\beqn
\cJ^i-Q\beta^i=Q v^i
&+&\alpha \sigma_{\rm c} \Big[w A^i_{~j}\cE^j + \epsilon^{ijk}u_j \cB_k \nonumber \\
&-&\alpha_{\rm d}\left(-w A^i_{~j}\cB^j + \epsilon^{ijk}u_j \cE_k\right)\Big],
\label{current}
\eeqn
with $A^i_{~j}:=\delta^i_{~j}-w^{-2}\bar u^i u_j$, $\bar
u^i=\gamma^{ij}u_j$, $u^\mu$ the four velocity, $v^i:=u^i/u^t$, and
$w=\alpha u^t$. $Q$ is a weighted charge density evaluated by $\pa_k
\cE^k /4\pi$.  $\sigma_{\rm c}$ is the conductivity
($\eta:=c^2/(4\pi\sigma_{\rm c})$ is the resistivity), and
$\alpha_{\rm d}$ denotes the so-called $\alpha$-parameter that
controls the mean-field dynamo~\cite{B2005,BDZ2013,Bugli,Tomei}. By
normalizing it with respect to the speed of light, we consider
$\alpha_{\rm d}$ as a dimensionless parameter in this paper.  In
contrast to the previous work~\cite{SFS2021}, we always consider the
cases of a non-zero value of $\alpha_{\rm d}$ in this paper.  $\kappa$
is chosen to be $10^5\,{\rm s}^{-1}$ for all the simulations.

In the present context, we suppose that $\sigma_{\rm c}$ and
$\alpha_{\rm d}$ should be determined by hypothetical turbulence 
motion of the fluid and resulting dynamo process. We
phenomenologically give these parameters from the consideration for
the plausible processes that play an important role in the remnant of
binary neutron star mergers as well as in the accretion disk around a
neutron star or a black hole.

\subsection{Choice of $\sigma_{\rm c}$ and $\alpha_{\rm d}$}\label{sec2-2}

In the presence of non-zero values of $\alpha_{\rm d}$ and
differential rotation, the so-called $\alpha$-$\Omega$ dynamo can be
activated.  In the non-relativistic case, the local analysis leads to
the following dispersion relation for the wave mode of the form
$\propto \exp(i\omega t -i k_i x^i)$~\cite{B2005}:
\beqn
(i\omega + \eta k^2)^2+i \alpha_{\rm d} S_\Omega k_\parallel c-\alpha_{\rm d}^2 k^2 c^2=0,
\label{disp}
\eeqn
where $k^2=k_i k^i$, $S_\Omega:=\pa \Omega/\pa \ln \varpi$, and
$k_\parallel$ denotes the wave number in the direction parallel to the
rotational axis.  To clarify the physical dimensions, $c$ is recovered
in this subsection.  The solution of Eq.~(\ref{disp}) is written as
\beqn
i\omega=-\eta k^2 \pm \sqrt{\alpha_{\rm d}^2 k^2 c^2 - i \alpha_{\rm d} S_\Omega k_\parallel c}.
\label{dispsol}
\eeqn

In this paper, we pay attention only to the case that a large-scale
dynamo plays a key role; i.e., we consider the case of
$\alpha_{\rm d}^2 k^2 c^2 \ll |\alpha_{\rm d} S_\Omega| k_\parallel c$.
For this case, Eq.~(\ref{dispsol}) is approximated as~\cite{B2005}
\beqn
i\omega \approx
-\eta k^2 \pm {1 - i \over \sqrt{2}}
\sqrt{|\alpha_{\rm d} S_\Omega| k_\parallel c},\label{dynamow}
\eeqn
and thus, the condition for the presence of the unstable modes 
becomes $\sqrt{|\alpha_{\rm d} S_\Omega| k_\parallel c /2}> \eta k^2$.
Focusing on the most optimistic case of $k^2=k_\parallel^2$, the
condition for $k$ to give an unstable mode becomes
\beqn
k &<& (8\pi^2 \sigma_{\rm c}^2|\alpha_{\rm d} S_\Omega|)^{1/3}c^{-1} \nonumber \\
&=&6.4\times 10^{-6}\,{\rm cm}^{-1}
\left({|\alpha_{\rm d}| \over 10^{-4}}\right)^{1/3}\nonumber \\
&& \times 
\left({\sigma_{\rm c} \over 3\times 10^{7}\,{\rm s}^{-1}}\right)^{2/3}
\left({|S_\Omega| \over 10^{3}\,{\rm rad/s}}\right)^{1/3},
\label{dynamok}
\eeqn
and the wave number of the fastest growing mode is 
\beqn
k_{\rm fast} &=& \left({\pi^2 \sigma_{\rm c}^2 |\alpha_{\rm d} S_\Omega| \over 2}
\right)^{1/3}c^{-1}
\nonumber \\ 
&=&2.5\times 10^{-6}\,{\rm cm}^{-1}
\left({|\alpha_{\rm d}| \over 10^{-4}}\right)^{1/3}
\nonumber \\
&& \times 
\left({\sigma_{\rm c} \over 3\times 10^{7}\,{\rm s}^{-1}}\right)^{2/3}
\left({|S_\Omega| \over 10^{3}\,{\rm rad/s}}\right)^{1/3}.
\label{fastD}
\eeqn
For the dynamo instability to take place in the remnant neutron star,
the typical scale of the unstable mode estimated broadly by $\sim
\pi/(2k_{\rm fast})$ should not be larger than the radius of the
neutron star $\sim 10$\,km. Thus, the condition to get the dynamo
instability in the neutron star becomes $\alpha_{\rm d}\sigma_{\rm
  c}^2 \agt 10^{10}\,{\rm s}^{-2}$.

For the fastest growing mode, the growth rate is written as
\beqn
\omega_{\rm max}&=&{3 \over 4}\left(
      {\pi \alpha_{\rm d}^2\sigma_{\rm c} S_\Omega^2 \over 4}\right)^{1/3}
=46\,{\rm s}^{-1}\left({|\alpha_{\rm d}| \over 10^{-4}}\right)^{2/3}
\nonumber \\
&& \times 
\left({\sigma_{\rm c} \over 3\times 10^{7}\,{\rm s}^{-1}}\right)^{1/3}
  \left({|S_\Omega| \over 10^{3}\,{\rm rad/s}}\right)^{2/3}. \label{growmax}
\eeqn
Thus, for massive neutron stars of mass $\sim 2.5M_\odot$ for which
$|S_\Omega|$ is typically of $O(10^3)\,{\rm rad/s}$, the growth
timescale of the electromagnetic fields is of order $10$\,ms.  For
accretion disks around a compact object of mass $\sim 3$--$10M_\odot$,
$|S_\Omega|$ can be slightly smaller, but for a compact orbit of radius
$\sim 100$\,km, the growth timescale is also as short as $\sim
10^2$\,ms.  For an accretion disk which orbits far from the central
object, $|S_\Omega|$ is smaller $\ll 10^3$\,rad/s, and the timescale
of the dynamo action is longer. Specifically, $|S_\Omega|$ is
approximately proportional to $R_{\rm disk}^{-3/2}$ where $R_{\rm
  disk}$ denotes the typical radius of the accretion disk, and thus,
$\omega_{\rm max}$ is approximately proportional to $R_{\rm
  disk}^{-1}$. This implies that for distant orbits, the dynamo action
becomes inefficient, if the values of $\sigma_{\rm c}$ and
$\alpha_{\rm d}$ for the non-compact disks are as large as those for
the compact disks.

Equation~(\ref{dynamow}) shows that in addition to the
growth (or damping) the electromagnetic field
oscillates with the angular frequency of
\beqn
\omega_{\rm osc}&:=&
\sqrt{{|\alpha_{\rm d} S_\Omega| k_\parallel c \over 2}}=61\,{\rm s}^{-1}
\left({|\alpha_{\rm d}| \over 10^{-4}}\right)^{1/2}
\nonumber \\
&\times& 
\left({|S_\Omega| \over 10^{3}\,{\rm rad/s}}\right)^{1/2}
\left({k_\parallel \over 2.5\times 10^{-6}\,{\rm cm}^{-1}}\right)^{1/2}.~~~
\eeqn
Thus, with our choice of $\sigma_{\rm c}$ and $\alpha_{\rm d}$ (see
below), the electromagnetic field changes the polarity with the period
of the order of $0.1$\,s($\sim 2\pi/\omega_{\rm osc}$) for the object
of total mass $\sim 3M_\odot$ and longer for the larger mass.


In the quasi-linear approximation under the assumption of the
isotropic turbulence~\cite{B2005}, the turbulent transport
coefficients, i.e., $\alpha_{\rm d}$ and $\eta$, are estimated by
\beqn
&&\alpha_{\rm d} \approx
-{1 \over 3c} \tau_{\rm cor} \langle u_i\omega^i\rangle,\\
&&\eta \approx {1 \over 3}\tau_{\rm cor} \langle u_i u^i\rangle,
\eeqn
where $\tau_{\rm cor}$ is a correlation time, $u_i$ is the fluctuation
part of the spatial velocity, and $\omega^i=\epsilon^{ijk}\pa_j u_k$:
the vorticity. $\langle \cdots \rangle$ denotes the ensemble averaging.
Assuming that $u_i$ and $\tau_{\rm cor}$ are comparable to the
Alfv\'en velocity and Alfv\'en timescale, the typical sizes for them
are evaluated by
\beqn
|u_i| &\approx &{B \over \sqrt{4\pi \rho}} 
= 2.0 \times 10^7\,{\rm cm/s}\left({B \over 10^{15}\,{\rm G}}\right) \nonumber \\
&&~~~~~\times 
\left({\rho \over 2 \times 10^{14}\,{\rm g\,cm^{-3}}}\right)^{-1/2},
\label{eq2.17}
\\
\tau_{\rm cor} & \approx & {R \over |u_i|}
= 50\,{\rm ms}\left( {R \over 10\,{\rm km}} \right)
\left({B \over 10^{15}\,{\rm G}}\right)^{-1}
\nonumber \\
&&~~~~~ \times 
\left({\rho \over 2 \times 10^{14}\,{\rm g\,cm^{-3}}}\right)^{1/2}, 
\label{eq2.18}
\eeqn
where $B$ is the typical magnetic-field strength, $\rho$ is the
rest-mass density, and $R$ is the radius of the neutron star.
Assuming that the order of magnitude of $\omega^i$ is the same as that
of $|u_i|/R$, we obtain $|\alpha_{\rm d}|=O(10^{-4})$ and
$\eta=O(10^{12})$--$O(10^{13})\,{\rm cm^2/s}$, i.e., $\sigma_{\rm
  c}=O(10^7)$--$O(10^8)\,{\rm s}^{-1}$. For these values, the
$\alpha$-$\Omega$ dynamo can be activated for long wavelength modes of
$\agt 1$\,km (cf.~Eq.~(\ref{dynamok})). In this paper, we broadly
suppose the situation for which the remnant neutron star and torus (or
disk) are unstable for the $\alpha$-$\Omega$ dynamo with these long
wavelength modes.

For dense accretion disks/tori surrounding a black hole/neutron star
which we consider in this paper, the typical sizes of $|u_i|$ and
$\tau_{\rm cor}$ are smaller and larger than those in
Eqs.~(\ref{eq2.17}) and (\ref{eq2.18}), respectively.  (We note that
$R$ should be replaced by the geometrical thickness of the disk/torus
which is $\alt 100$\,km.)  However, the order of the magnitude is not
significantly different from those for the neutron star. Hence, we
employ the same values of $\sigma_{\rm c}$ and $\alpha_{\rm d}$ both
for the remnant neutron star and accretion disks surrounding the
central compact objects for simplicity, while we perform several
simulations varying these parameters for a certain range.

We note that in the late evolution stage of the remnant neutron star,
the degree of the differential rotation is likely to become weak due
to the MHD effects (cf.~Sec.~\ref{sec4}). Even for such a state, the
magnetic-field amplification may be still preserved by the
$\alpha$-dynamo, for which the necessary condition (from
Eq.~(\ref{dispsol}) with $S_\Omega=0$) is written as~\cite{B2005} 
\beq
k < {4\pi\sigma_{\rm c}|\alpha_{\rm d}| \over c},
\eeq
or equivalently
\beq
\lambda={2\pi \over k} > 50\,{\rm km}
\left({|\alpha_{\rm d}| \over 10^{-4}}\right)^{-1}
\left({\sigma_{\rm c} \over 3\times 10^{7}\,{\rm s}^{-1}}\right)^{-1}. 
\eeq
Thus, with the setting of $\alpha_{\rm d}=10^{-4}$ and $\sigma_{\rm c}
\alt 10^8\,{\rm s}^{-1}$, the wavelength for the unstable modes is so
long that the effect of the $\alpha$-dynamo is minor.  Hence, the
magnetic field is likely to decay with the time scale of $\agt 4\pi
\sigma_{\rm c}/(kc)^2$ where $\sim k^{-1}$ denotes the curvature scale
of the magnetic-field lines.  In reality, in the absence of the
differential rotation, a convection or turbulence is not likely to be
preserved, and thus, supposing the mean-field dynamo is unlikely to be
correct.  Thus in this paper, we do not touch on the very late time
evolution of the system, in which the magnetic-field decays due to the
resistivity.

Finally we note that $\alpha_{\rm d}$ is not a pure scalar but an
axial scalar because it has the same polarity as that of the toroidal
magnetic field. In the simulation of this paper we assume the plane
symmetry with respect to the equatorial plane. Thus, $\alpha_{\rm d}$
should have the reflection anti-symmetry for the change of $z
\rightarrow -z$, and thus, $\alpha_{\rm d}=0$ on the $z=0$
(equatorial) plane. To impose this condition, we employ the functional
form of $\alpha_{\rm d}$ as
\beq
\alpha_{\rm d}=\alpha_{\rm d,0}\left[
{2 \over \exp(-|z|/z_{\rm c}) + 1}-1  
  \right]{z \over |z|}, 
\eeq
where we choose $z_{\rm c}=0.5$\,km in this paper, and $\alpha_{\rm
  d,0}$ is a constant chosen based on the estimate for the approximate
magnitude of $\alpha_{\rm d}$ shown above (hereafter we denote
$\alpha_{\rm d,0}$ simply by $\alpha_{\rm d}$).  Since the value of
$z_{\rm c}$ is much smaller than the geometrical thickness of the disk
and radius of the neutron star, the dynamo effect is assumed to be
present for a wide region in which the matter is present.

\subsection{Diagnostics}

We always calculate the following quantities for the simulation
results: average specific entropy $\langle s \rangle$ and average
electron fraction $\langle Y_e \rangle$ both for the matter located
outside the black hole and for the ejecta (see the method for
identifying the ejecta below). For the former case, these average
quantities are defined by
\beqn
\langle s \rangle &:=&{1 \over M_{\rm mat}} \int_{\rm out} \rho_* s \,d^3x, \\
\langle Y_e \rangle &:=&{1 \over M_{\rm mat}} \int_{\rm out} \rho_* Y_e \,d^3x, 
\eeqn
where $M_{\rm mat}$ denotes the rest mass of the matter located
outside the black hole, defined by
\beqn
M_{\rm mat}&:=& \int_{\rm out} \rho_* \,d^3x, 
\eeqn
and $\int_{\rm out}$ implies that the volume integral is performed for
the matter located outside the black hole. For the ejecta component,
the volume integral is performed for the matter that satisfies the
ejecta criterion (see below).

The kinetic energy and the electromagnetic energy of the system are
defined by
\footnote{In our previous paper~\cite{SFS2021}, all the numerical
  results for $E_{\rm B}$ were twice larger than the correct values
  because the factor of $8\pi$ was written as $4\pi$ in the code by
  typos. 
}
\beqn
E_{\rm kin}&:=&{1 \over 2} \int_{\rm out} \rho_* h u_i v^i d^3x,\\
E_{\rm B}&:=&{1 \over 8\pi} \int_{\rm out} (B^2 + E^2) \sqrt{-g}\,d^3x.
\eeqn

The ejecta component is determined using the same criterion as in
Refs.~\cite{Fujiba20,Fujiba2020}; we identify a matter component with
$|h u_t| > h_{\rm min}$ located in a far region as the ejecta. Here
$h_{\rm min}$ denotes the minimum value of the specific enthalpy in
the adopted equation-of-state table, which is $\approx 0.9987c^2$.
For the matter escaping from a sphere of its radius $r=r_{\rm ext}$,
we define the ejection rates of the rest mass and energy (kinetic
energy plus internal energy) at a given radius and time by
\beqn
\dot M_{\rm eje}&:=&\oint \rho \sqrt{-g} u^i dS_i, \label{ejectrate} \\
\dot E_{\rm eje}&:=&\oint \rho \hat e \sqrt{-g} u^i dS_i,
\label{ejectrate2}
\eeqn
where $\hat e:=h \alpha u^t-P/(\rho \alpha u^t)$.  The surface
integral is performed at $r=r_{\rm ext}$ with $dS_i=\delta_{ir}r_{\rm
  ext}^2 \sin\theta d\theta d\varphi$ for the ejecta
component. $r_{\rm ext}$ is chosen to be $\approx 6000$\,km for black
hole-disk systems (with the mass of the black hole of $10M_\odot$) and
1500\,km for neutron star-torus systems in this work.


As in our previous paper~\cite{SFS2021}, the total rest mass and
energy (excluding the gravitational potential energy and
electromagnetic energy) of the ejecta (which escape away from a sphere
of $r=r_{\rm ext}$) are calculated, respectively, by
\beqn
M_{\rm eje}(t)&:=&\int^t \dot M_{\rm eje} dt, 
\\
E_{\rm eje}(t)&:=&\int^t \dot E_{\rm eje} dt. 
\eeqn
Far from the central object, $E_{\rm eje}$ is approximated by the
sum of the rest-mass energy and kinetic energy, 
as we discussed in Ref.~\cite{SFS2021}.
Since the ejecta velocity can be relativistic in this work, we first
define an average Lorentz factor of the ejecta (for the component that
escapes from a sphere of $r=r_{\rm ext}$) by
\beq
\Gamma_{\rm eje}:=\sqrt{{E_{\rm eje}-GM M_{\rm
    eje}/r_{\rm ext} \over M_{\rm eje}c^2}}, \label{eq230}
\eeq
where $M$ denotes the total gravitational mass of the system and $G$
is recovered to clarify that the last term in the numerator of
Eq.~(\ref{eq230}) approximately denotes the gravitational potential
energy of the matter at $r=r_{\rm ext}$. Finally, the average ejecta
velocity is calculated by $v_{\rm eje}:=\sqrt{1-\Gamma_{\rm eje}^{-2}}$. 

\section{Evolution of black hole-disk systems}\label{sec3}

\subsection{Setup}\label{sec3-1}

First we evolve systems composed of a spinning black hole and a disk
surrounding it. We employ the same initial conditions as models M10L05
and M10H05 of Ref.~\cite{Fujiba20b}, which are in equilibrium states
in the absence of MHD effects. Specifically, the initial conditions
for models M10L05 and M10H05 are composed of a black hole of the
initial mass $M_{\rm BH,0}=10M_\odot$ and dimensionless spin $\chi
\approx 0.8$ and of a disk of mass $0.1M_\odot$ and $3M_\odot$,
respectively. The reason that we choose the $10M_\odot$ black hole
(rather than the lower-mass ones) is that with a high-mass black hole,
the finest grid spacing near the black hole can be taken to be large,
i.e., the time step determined by the minimum grid spacing, $\Delta
x_0$, can be large, and hence, the computational costs are saved for
the fixed simulation time of $\sim 5$\,s. Although the black-hole mass
employed is larger than the neutron-star mass, the disk width
(determined by the location of the outer edge of the disk) is $\sim
200$\,km which is as large as that for the torus formed around the
neutron star after binary neutron-star mergers. Thus, the accretion
disk has a structure similar to the torus that surrounds the remnant
neutron star of binary-neutron-star mergers.

Models M10L05 and M10H05 were already evolved in viscous hydrodynamics
with plausible values of the (viscous) $\alpha$-parameter of 0.05 and
the scale height of $2M_{\rm BH,0}(\approx 30$\,km) in a previous
paper~\cite{Fujiba20b}.  We compare the results obtained in the
present MHD simulations with those in the viscous hydrodynamics ones
for several choices of $\sigma_{\rm c}$ and $\alpha_{\rm d}$, and show
that the results by these two different approaches provide
quantitatively similar results (in particular for the low-mass disk
models).

\begin{table}[t]
  \caption{Initial conditions and setup for the numerical simulations
    of a spinning black hole of mass $M_{\rm BH,0}\approx 10M_\odot$
    and disks of mass $M_{\rm disk}\approx 0.1M_\odot$ or
    $3.0M_\odot$.  For all the models the dimensionless spin of the
    black hole is $\chi \approx 0.8$ (see also Ref.~\cite{Fujiba20b}).
    For the low-mass disk models, $E_{\rm B}=3.5\times 10^{46}$\,erg
    and $E_{\rm kin}=1.4\times 10^{52}$\,erg initially. For the
    high-mass disk models, $E_{\rm B}=1.1\times 10^{47}$\,erg and
    $E_{\rm kin}=3.7\times 10^{53}$\,erg initially. For the
    resolution, M and H denote the medium and high resolutions,
    respectively (see the text for more details).  }
\begin{tabular}{ccccc} \hline
  ~~Model~~ & ~$M_{\rm disk}/M_\odot$~ 
  & ~~$\sigma_{\rm c}\,{\rm (s^{-1})}$~~ &~$\alpha_{\rm d}$~ & Resolution\\
 \hline \hline
M10L80  & 0.1 & $1 \times 10^{8}$ & $1 \times 10^{-4}$ & M, H \\
M10L75a & 0.1 & $3 \times 10^{7}$ & $1 \times 10^{-4}$ & M, H \\
M10L75b & 0.1 & $3 \times 10^{7}$ & $2 \times 10^{-4}$ & M, H \\
M10L70 & 0.1 & $1 \times 10^{7}$ & $1 \times 10^{-4}$ & H \\
M10H80  & 3.0 & $1 \times 10^{8}$ & $1 \times 10^{-4}$ & M \\
M10H75a & 3.0 & $3 \times 10^{7}$ & $1 \times 10^{-4}$ & M \\
M10H75b & 3.0 & $3 \times 10^{7}$ & $2 \times 10^{-4}$ & M \\
M10H70  & 3.0 & $1 \times 10^{7}$ & $1 \times 10^{-4}$ & M \\
 \hline
\end{tabular}
\label{table1}
\end{table}

For the present MHD simulations, we initially superimpose a purely
toroidal magnetic field in a high-density region of the disk as
\beqn
\cB^T=\varpi \cB^\varphi
=A_0 z \,{\rm max}\left({P \over P_{\rm max}} - 0.04, 0\right),
\label{initoro}
\eeqn
where $P$ is the gas pressure and $P_{\rm max}$ is the maximum value
of $P$. The poloidal component of $\cB^i$ is set to be zero initially
and the electric field is determined by the ideal MHD condition of
$\cE^i=-\epsilon^{ijk}u_j\cB_k/w$ for simplicity.  The dependence on
the $z$ coordinate in Eq.~(\ref{initoro}) stems from the reflection
anti-symmetry for $\cB^T$ with respect to the $z=0$ plane.  $A_0$ is a
constant, and in this work, we choose it so that the electromagnetic
energy is $E_{\rm B} \approx 3.5\times 10^{46}$\,erg for the low-mass
disk models and $E_{\rm B} \approx 1.1\times 10^{47}$\,erg for the
high-mass disk models. For both cases, the initial values of $E_{\rm
  B}$ is much smaller than the internal energy and rotational kinetic
energy of the system. Because the magnetic-field strength increases
exponentially with time until the saturation of the growth universally
in the presence of the mean-field dynamo, the final result does not
depend essentially on the initial field strength.

In the absence of the dynamo term ($\alpha_{\rm d}=0$) and in axial
symmetry with this type of the initial condition, the magnetic field
of a purely toroidal field should be simply preserved or decay with
the resistive timescale determined by $\sigma_{\rm c}$~(see, e.g.,
appendix B of Ref.~\cite{SFS2021}).  On the other hand, in the
presence of the dynamo term, the poloidal field is generated from the
toroidal field, and subsequently, due to the $\alpha$-$\Omega$ dynamo
effect, winding, and the MRI, the strength of both the toroidal and
poloidal fields is enhanced. Note that the early magnetic-field growth
is driven purely by the $\alpha$-$\Omega$ dynamo effect for the
present initial condition only with the toroidal magnetic field in the
axisymmetric simulation.

Table~\ref{table1} summarizes the models which we consider in this
paper.  The values of $\sigma_{\rm c}$ and $\alpha_{\rm d}$ are chosen
so that long-wavelength dynamo modes become unstable as discussed in
Sec.~\ref{sec2-2}. We note that for a high value of $\sigma_{\rm c} >
10^8\,{\rm s}^{-1}$ with $\alpha_{\rm d}=10^{-4}$, the amplification
of the magnetic field by the $\alpha$-$\Omega$ dynamo proceeds
initially to an extremely high level perhaps due to the amplification
in the shorter-wavelength modes.  Because it is not clear to us
whether such an extreme amplification is realistic or not, we do not
pay attention to such cases in this paper. A significant
amplification of the magnetic-field strength of long-wavelength modes
and resulting effects are induced even with $\sigma_{\rm c} =
10^7$--$10^8\,{\rm s}^{-1}$.

Following our previous work~\cite{Fujiba20,Fujiba20b}, we employ a
nonuniform grid for the two-dimensional (cylindrical) coordinates $(x,
z)$ in the simulation: For $x \leq x_0=0.9GM_{\rm BH,0}/c^2$, a uniform
grid is used with the grid spacing $\Delta x_0=0.015GM_{\rm
  BH,0}/c^2(\approx 0.22$\,km in this work), and for $x > x_0$, $\Delta
x$ is increased uniformly as $\Delta x_{i+1}=\eta_\Delta \Delta x_i$
where the subscript $i$ denotes the $i$-th grid with $x=0$ at
$i=0$. For $z$, the same grid structure as for $x$ is used for all the
models.  We refer to the grid resolution with $\eta_\Delta=1.01$ and
1.008 as medium (M) and high (H) resolutions. The two grid resolutions
are employed to confirm the reasonable convergence of the numerical
result for low-disk mass models (M10L80, M10L75a, and M10L75b).  The
black-hole horizon is always located in the uniform grid zone.  The
location of the outer boundaries along each axis, $L$, is $\approx
1.1\times 10^4$\,km irrespective of the models.

\subsection{Numerical results}\label{sec3-2}

\begin{figure*}[t]
\includegraphics[width=80mm]{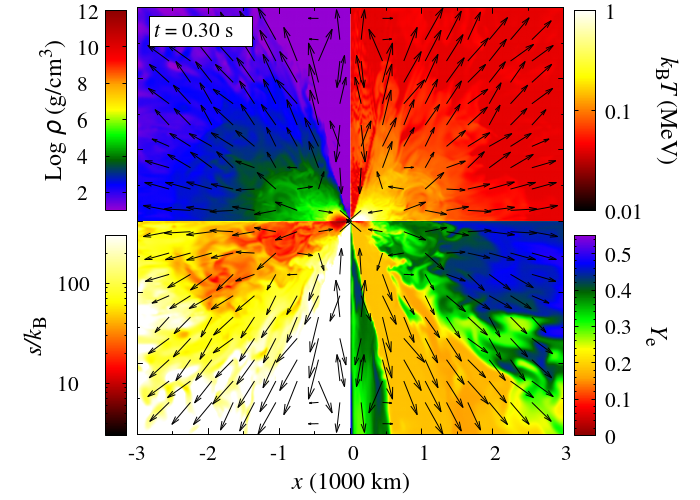}~
\includegraphics[width=80mm]{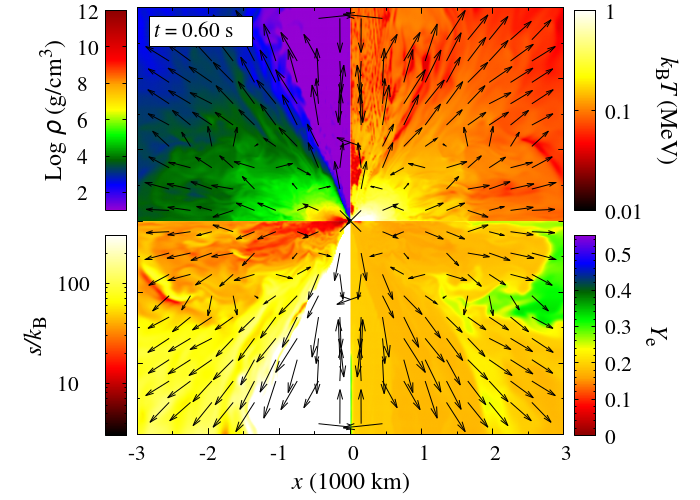} \\
\includegraphics[width=80mm]{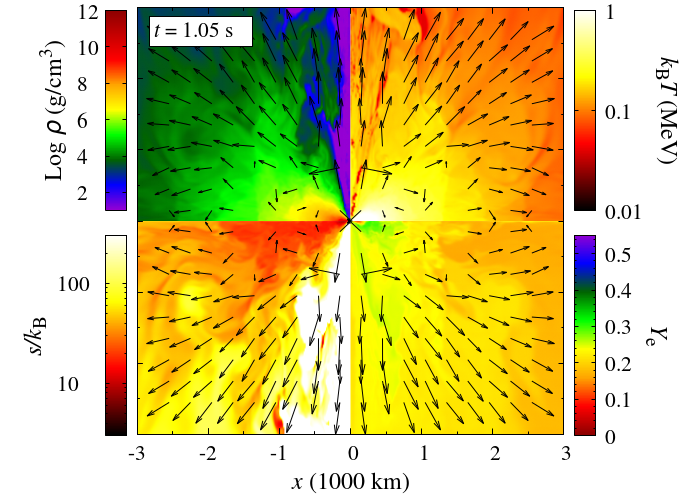}~
\includegraphics[width=80mm]{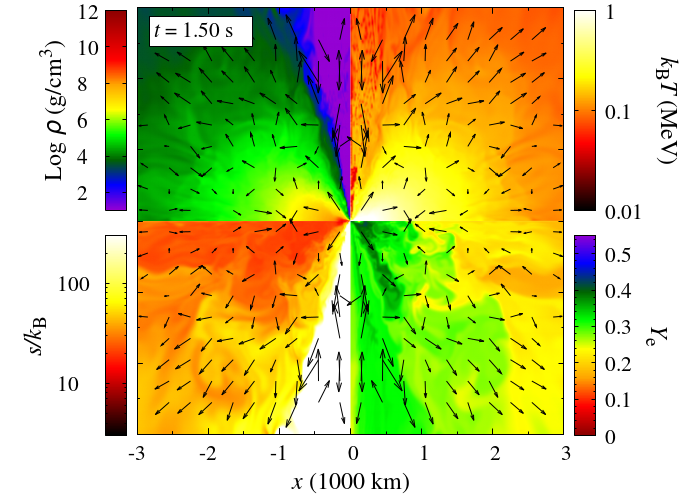} \\
\includegraphics[width=80mm]{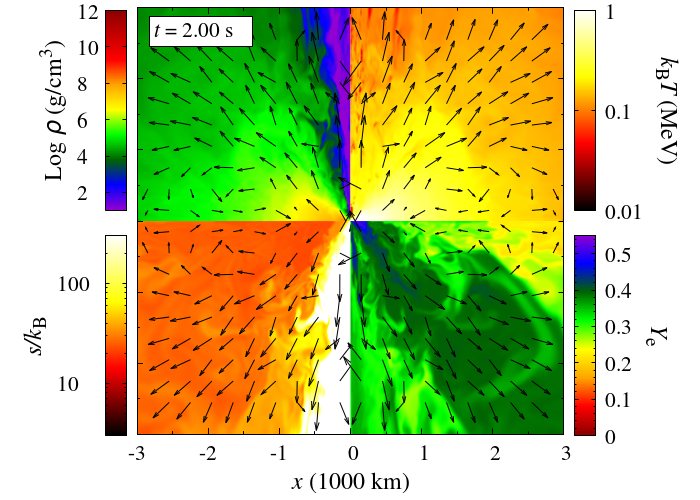}~
\includegraphics[width=80mm]{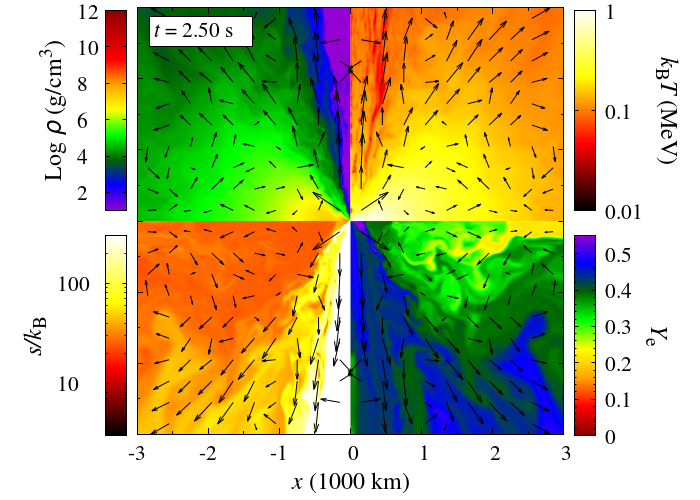} 
  \caption{Snapshots of the rest-mass density in units of ${\rm
      g/cm}^3$, temperature ($k_{\rm B}T$ in units of MeV with $k_{\rm B}$ being
    the Boltzmann constant), specific entropy $s$ in units of $k_{\rm B}$,
    and electron fraction $Y_e$ at selected time slices for model
    M10L75a with the high-resolution run. The arrows denote the
    velocity field of $(v^x, v^z)$.
\label{fig1}}
\end{figure*}
\begin{figure*}[t]
\includegraphics[width=175mm]{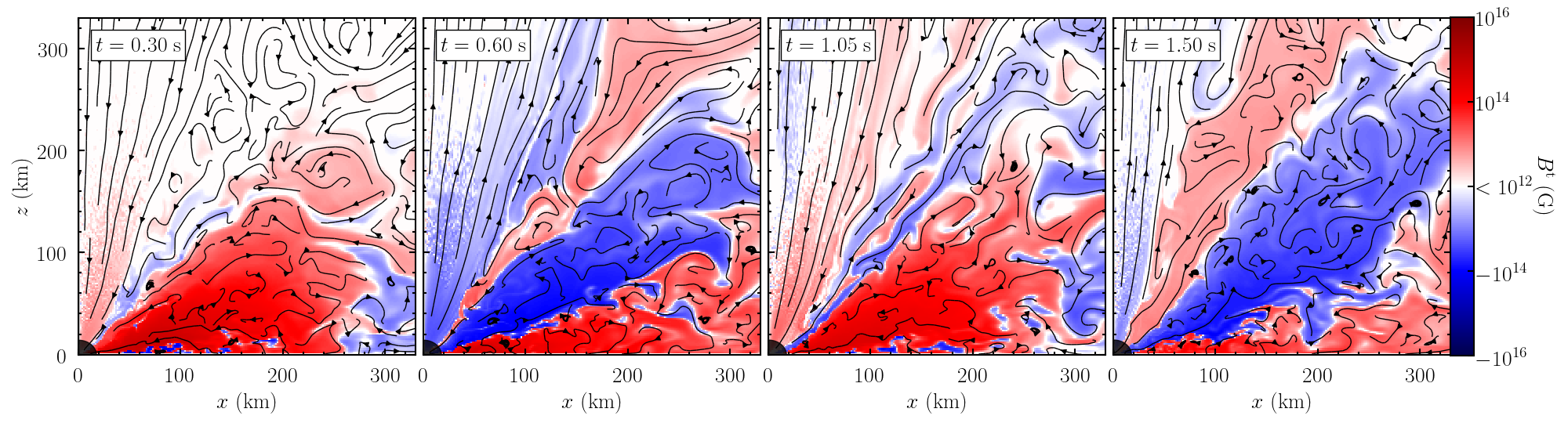}
  \caption{Evolution of the poloidal magnetic-field lines 
    together with the toroidal magnetic-field strength (color profile)
    in the near zone of $|x| \leq 330$\,km and $|z| \leq 330$\,km  
    for model M10L75a with the high-resolution run. 
\label{fig2}}
\end{figure*}

\subsubsection{Evolution of the system}

First, we briefly summarize the evolution process of the disk
determined by the dynamo action described in Sec.~\ref{sec2-2}.  By
the $\alpha$-$\Omega$ dynamo effect, the poloidal magnetic field is
developed from the purely toroidal field initially given.  By the
subsequent $\alpha$-$\Omega$ dynamo effect and the winding of the
magnetic field lines of the generated poloidal field, the toroidal
field is also enhanced. After the significant amplification of the
magnetic-field strength, with which its fastest growing mode can be
resolved, the MRI also plays an important role~\cite{BH98}. Because of
the resulting development of the turbulent state in the disk, the disk
matter expands and a part of it is ejected from the disk in a
quasi-spherical manner: The major part is ejected primarily toward the
non-polar direction because of the presence of the angular momentum
barrier, but an outflow of the low-density matter is also observed in
the polar direction (see Fig.~\ref{fig1} for model M10L75a). A funnel
with the half opening angle of $10^\circ$--$15^\circ$ is typically
formed around the $z$-axis after the outflow is steadily driven.  The
resultant funnel region has a high ratio of the magnetic pressure to
the gas pressure (i.e., low-$\beta$ plasma).  All these features are
found irrespective of the values of $\sigma_{\rm c}$ and $\alpha_{\rm
  d}$ employed in the present work, and are qualitatively very similar
to those found in viscous hydrodynamics
simulations~\cite{Fujiba20,Fujiba20b}. However, the qualitative and
quantitative details are different between the results of viscous
hydrodynamics and MHD (see below). Also the MHD results depend
quantitatively on the values of $\sigma_{\rm c}$ and $\alpha_{\rm d}$.


The matter outflow is accompanied by the ejection of the magnetic
loops from the disk. This generates the large-scale poloidal and
toroidal magnetic fields outside the disk.  At the same time, in the
disk, highly disturbed magnetic fields are developed reflecting its
turbulent motion. Furthermore, in the presence of the dynamo, the
polarity of the magnetic field is often changed.  Figure~\ref{fig2}
displays the poloidal magnetic field lines together with the toroidal
magnetic-field strength for model M10L75a. Here, the red and blue
colors denote that the toroidal field is positive and negative,
respectively. This figure shows that highly distorted
poloidal-magnetic fields are indeed developed in the accretion disks
after the enhancement of the magnetic-field strength by the dynamo
process reflecting the development of a turbulent motion inside the
disk. Some of the field lines are extended outside the disk and form
the global magnetic fields.  It is also found that an aligned magnetic
field is developed in the funnel region near the $z$-axis. All these
features in the outcome are qualitatively the same as those often
found in the ideal MHD simulations for the accretion disks around the
black hole
(e.g.,~Refs.~\cite{FTQFK19,LTQ,Gammie,DeV,Hirose,MG04,MTB12,MTSR14}),
for which the ideal MHD simulations are for most cases started from a seed
poloidal field and the turbulence is developed purely (in the first
principle manner) by the MRI.

One interesting feature in the simulations with the mean-field dynamo
term is that the magnetic-field polarity changes~\cite{Sadowski2015}
in a quasi-periodic manner, with the period of several hundred ms in
our present setting (see Sec.~\ref{sec2-2}).  The color profile of
Fig.~\ref{fig2} illustrates that the toroidal-field polarity indeed
changes as found in the high-resolution three-dimensional ideal MHD
simulation~\cite{FTQFK19,LTQ}.  Also, near the $z$-axis, the
poloidal-field polarity changes in the same period as that for the
toroidal field (see the arrows of the poloidal field lines in
Fig.~\ref{fig2}).  This implies that an entirely coherent, aligned
magnetic field is not established in the large scale, although
coherent magnetic-field lines near the $z$-axis are locally developed,
and that during the change of the polarity of the poloidal field near
the $z$-axis, not an aligned magnetic field but a disturbed field
configuration transiently appears.

In the late evolution stage in which the mass and density of the
accretion disk decrease, the magnetic-field strength decreases. In
particular in the very late stage in which the matter motion is
dominated by the turbulent motion (not by a coherent rotational motion
in the disk), the poloidal field near the $z$-axis does not have a
very aligned structure (and as a result, the polar outflow
ceases). All these features are found irrespective of the mean-field
dynamo parameters employed.

\begin{figure*}[t]
\includegraphics[width=84mm]{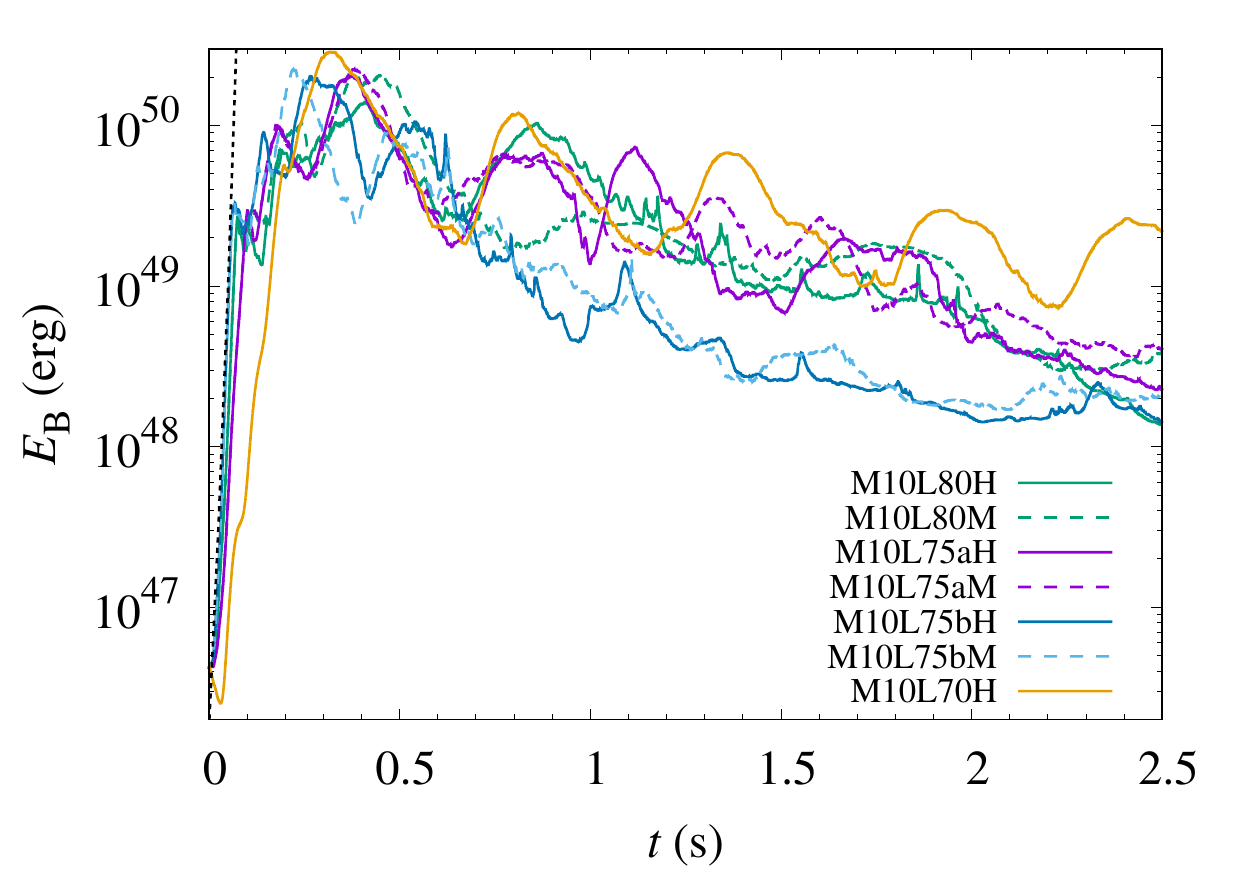}~~
\includegraphics[width=84mm]{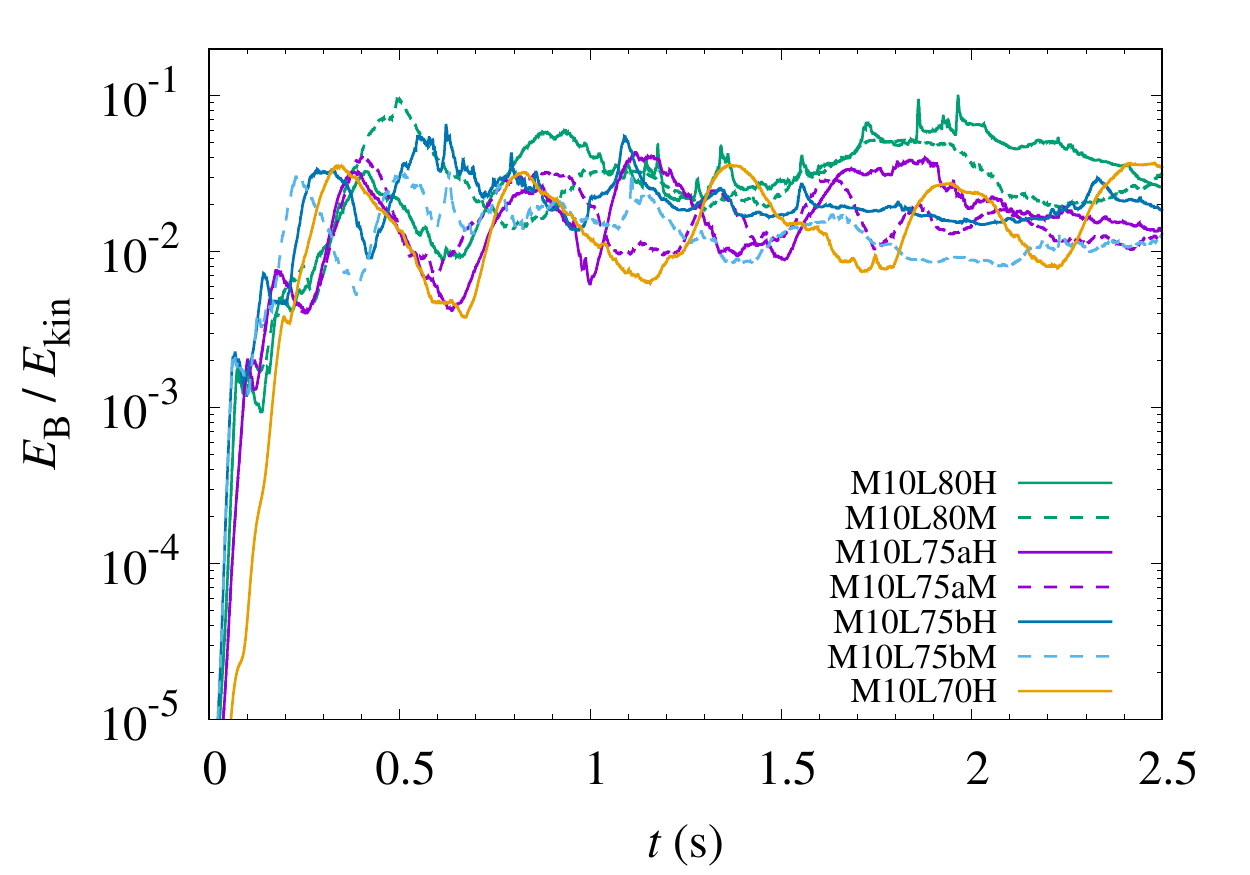} \\
\includegraphics[width=84mm]{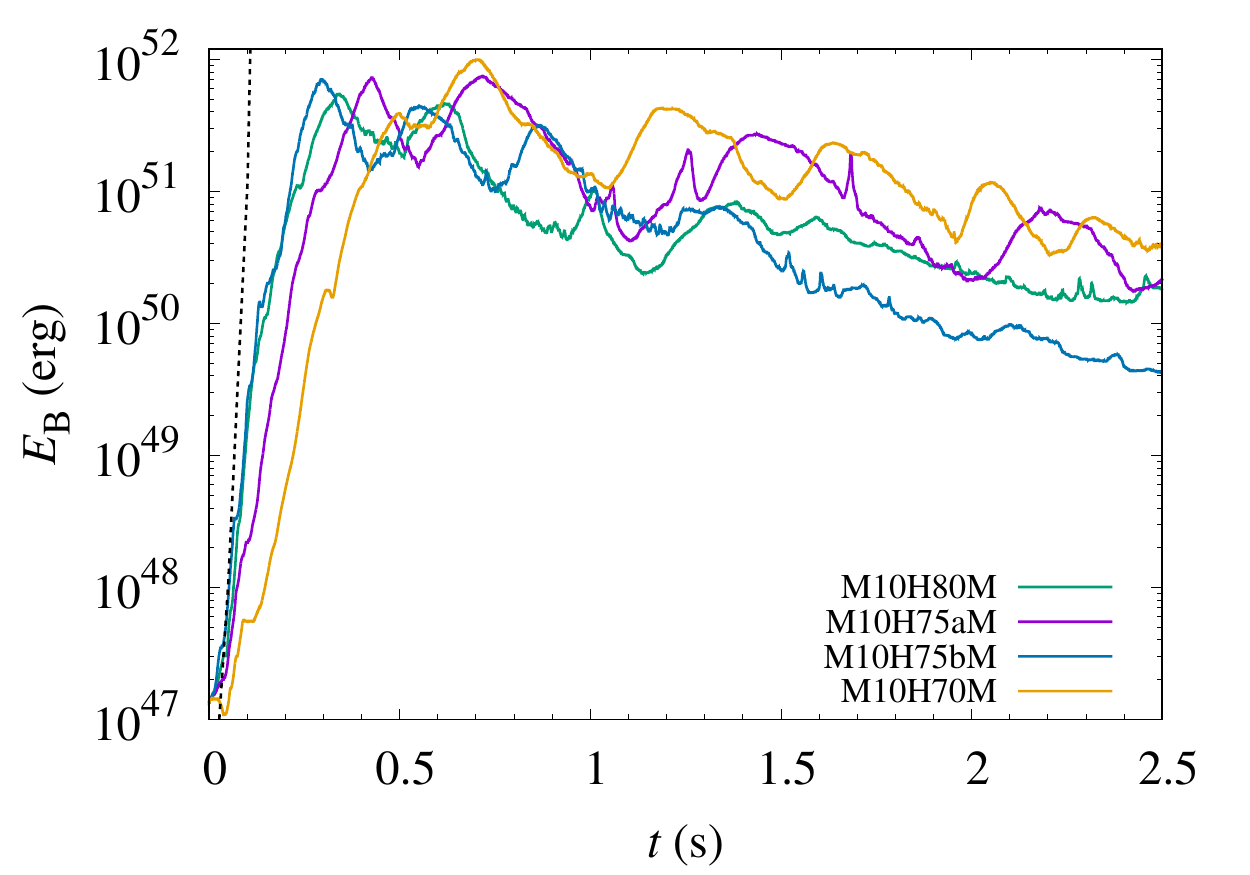}~~
\includegraphics[width=84mm]{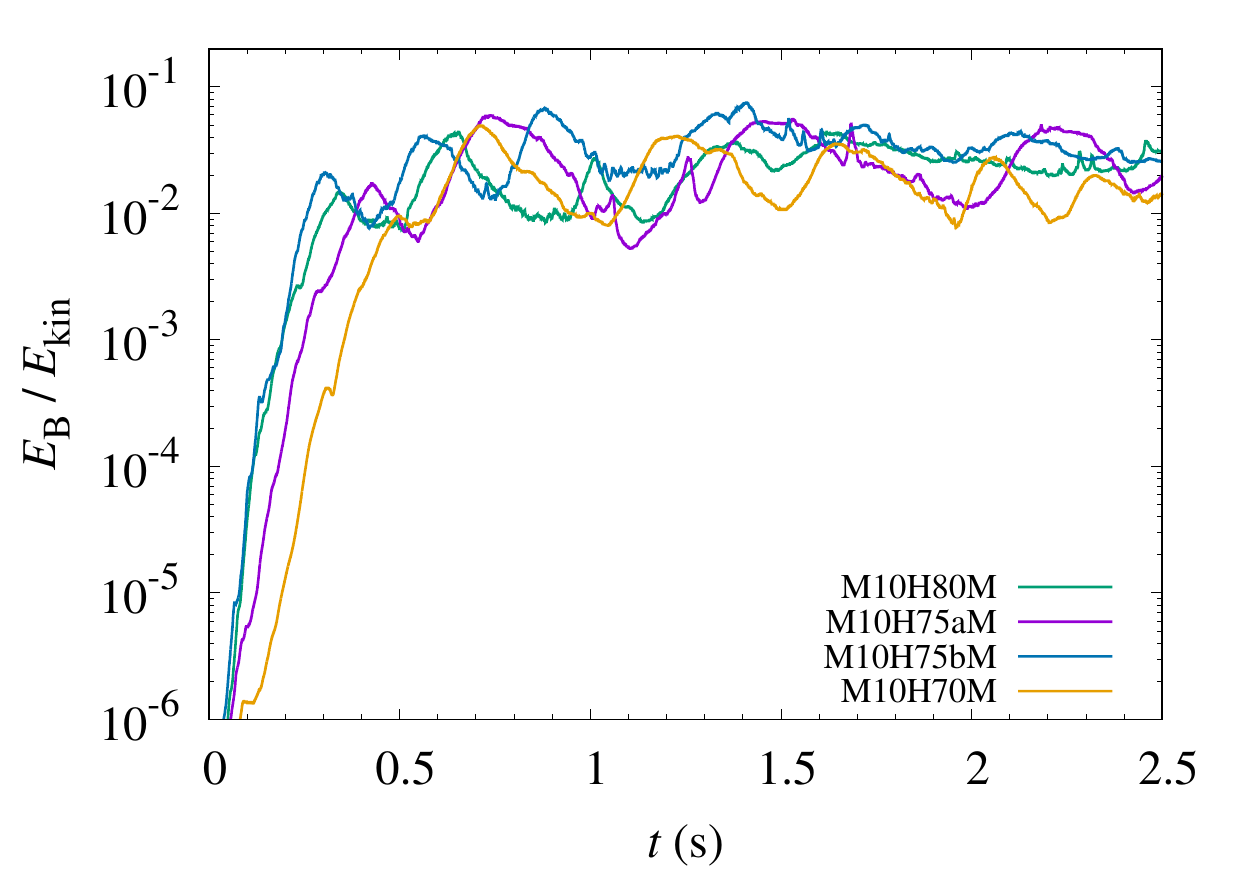}
\caption{Upper panels: Evolution of the electromagnetic energy and
  ratio of the electromagnetic energy to the kinetic energy for the
  systems of a black hole and a low-mass disk listed in
  Table~\ref{table1}.  Lower panels: The same as the upper panels but
  for the systems of a black hole and a high-mass disk listed in
  Table~\ref{table1}.  The black dotted lines in the left two panels show
  $\propto \exp(2\omega_{\rm max}t)$ with $\alpha_{\rm d}=10^{-4}$,
  $\sigma_{\rm c}=10^8\,{\rm s}^{-1}$, and $|S_\Omega|=10^3\,{\rm rad/s}$:
  cf.~Eq.~(\ref{growmax}).
\label{fig3}}
\end{figure*}
\begin{figure*}[t]
(a)\includegraphics[width=84mm]{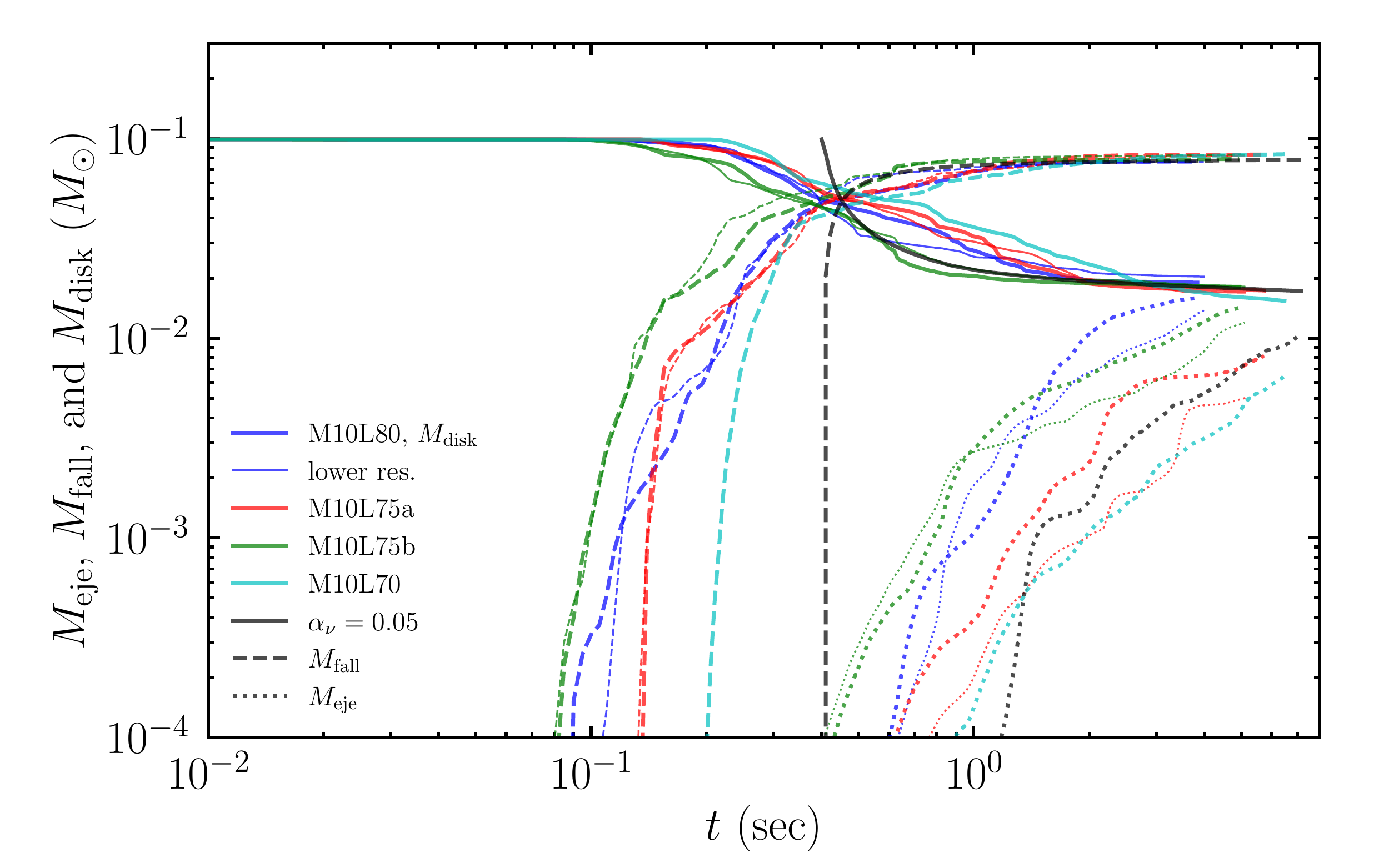}
(b)\includegraphics[width=84mm]{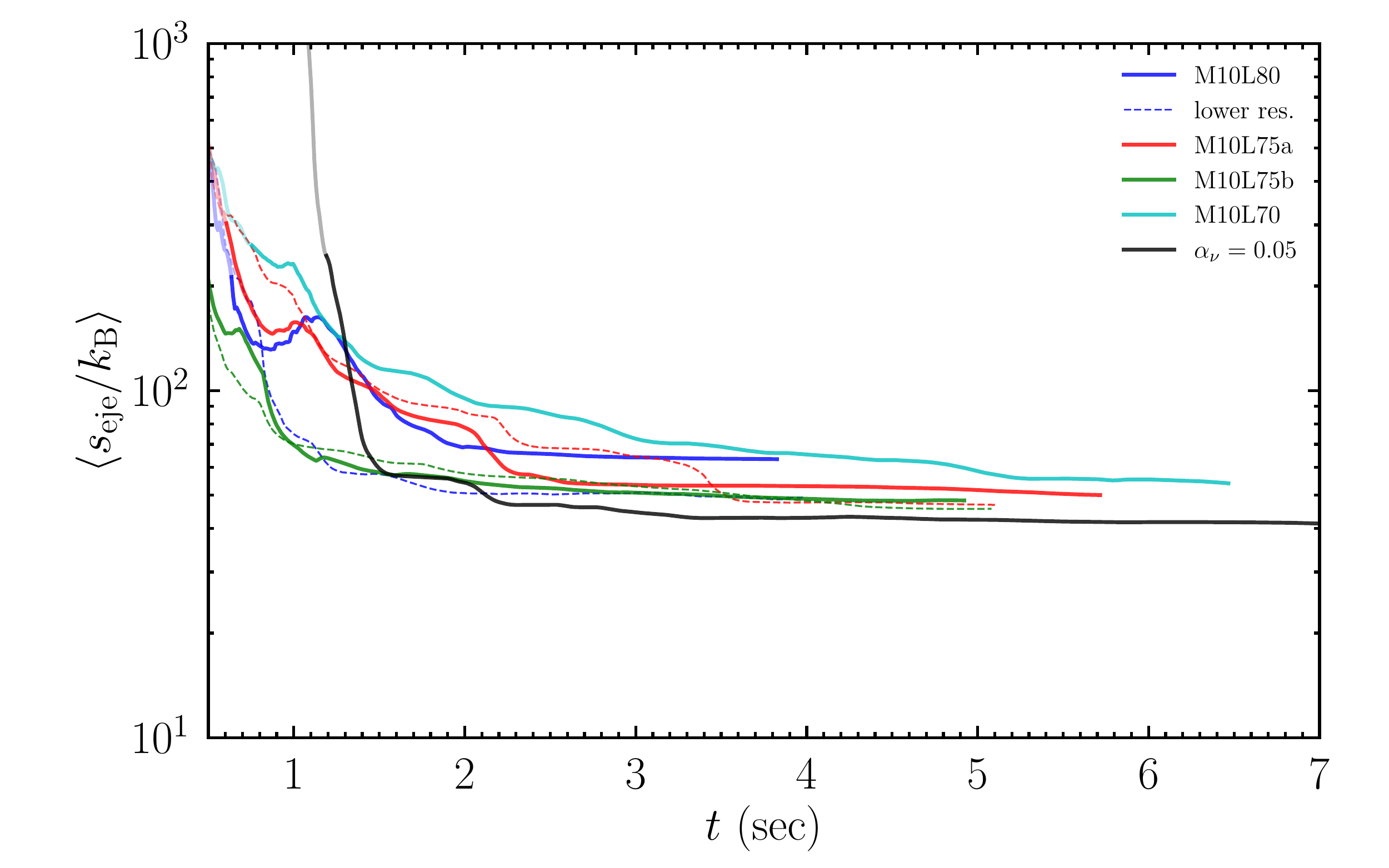}\\
(c)\includegraphics[width=84mm]{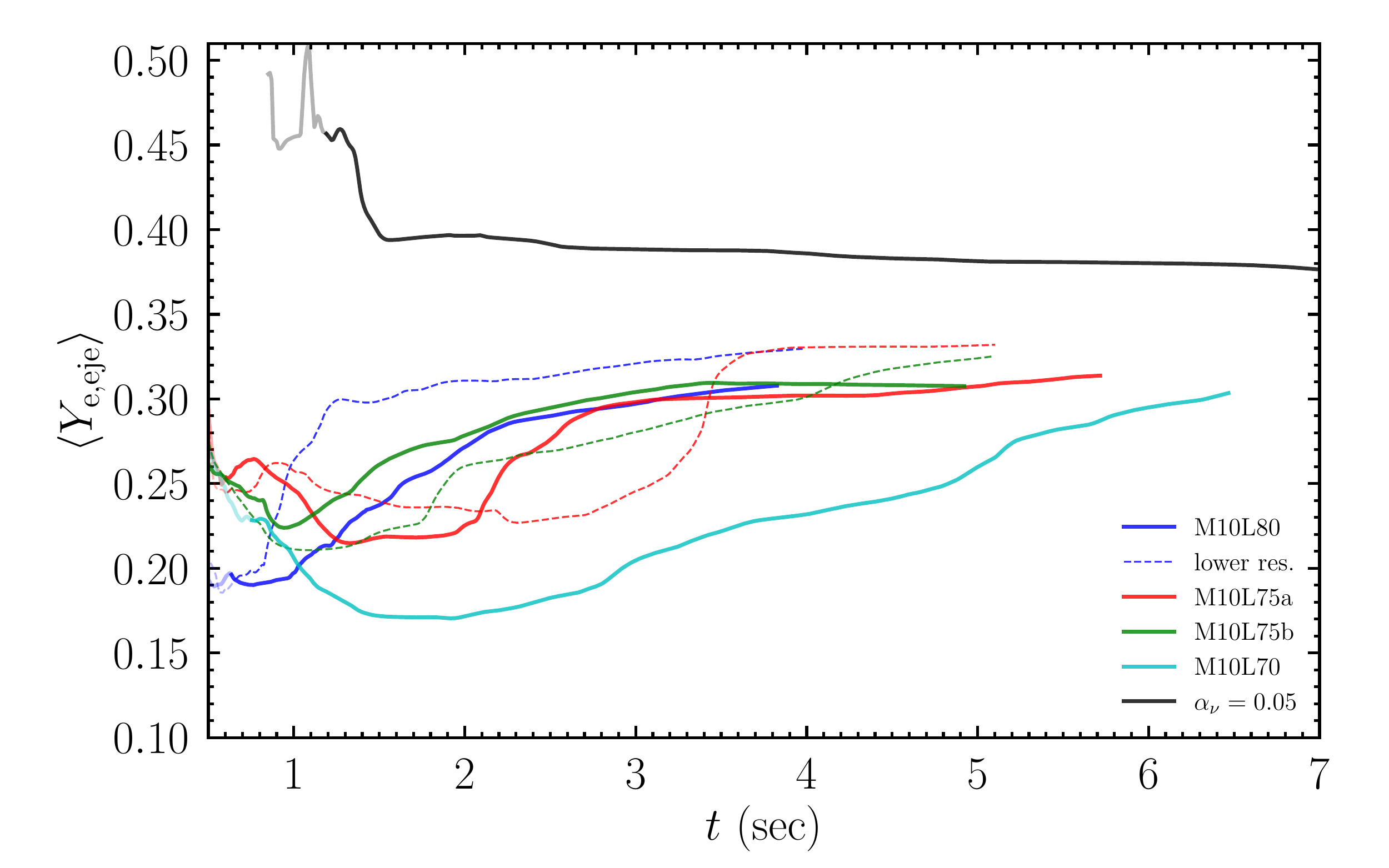}
(d)\includegraphics[width=84mm]{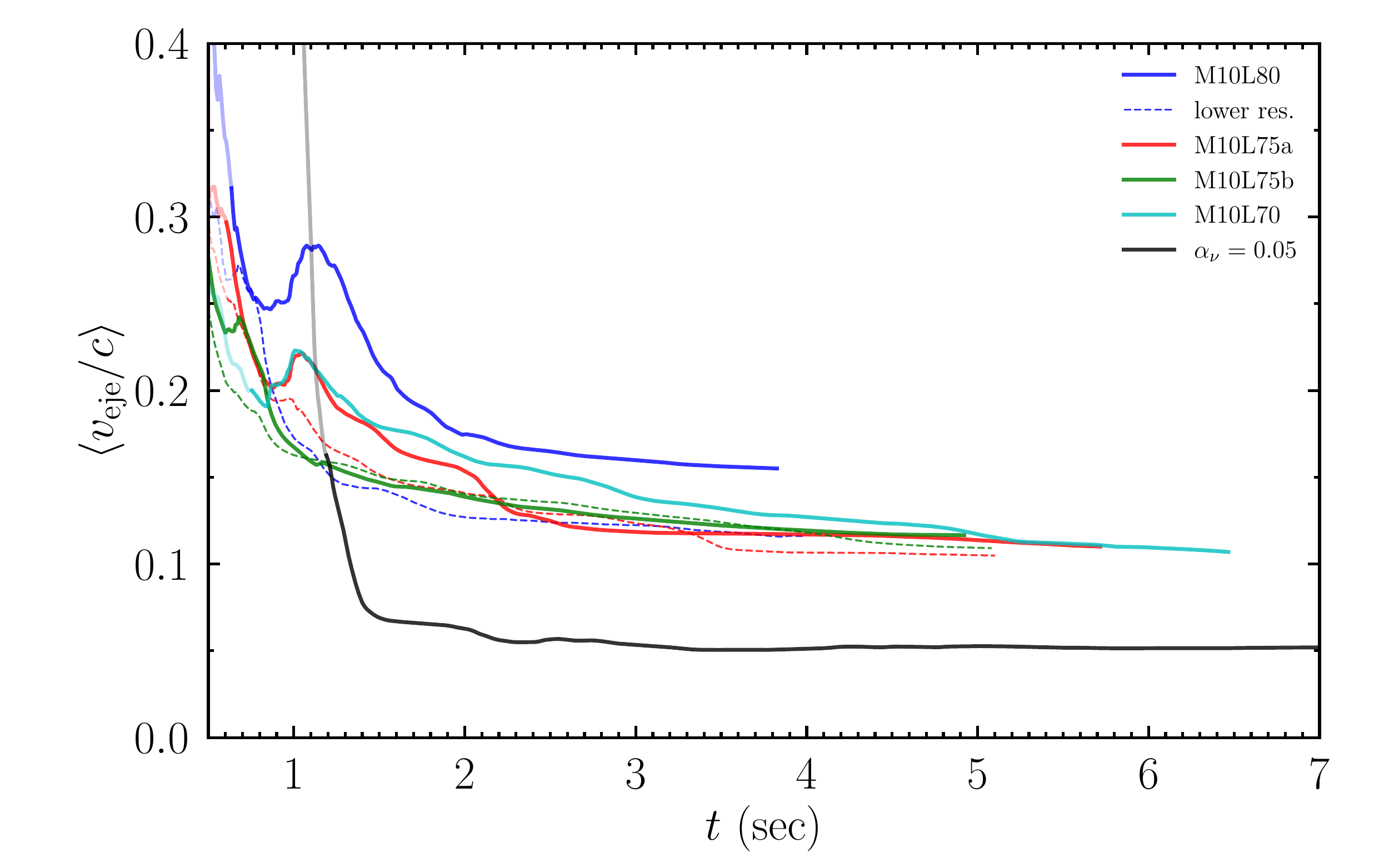}
  \caption{Evolution of several quantities for the low-mass disk
    surrounding a black hole. (a) Mass of the matter located outside
    the black hole (solid curves), that falls into the black hole
    (dashed curves), and ejected from the system (dashed curves).
    The thick and thin curves denote the results with the
    high- and medium-resolution runs, respectively. 
    (b) the average entropy per baryon for the
    ejecta;  (c) the average value of $Y_e$ for the ejecta; and (d) the
    average velocity for the ejecta. For (b)--(d), the solid and
    dashed curves denote the high- and medium-resolution results. For
    comparison, the results by a viscous hydrodynamics simulation
    performed in Ref.~\cite{Fujiba20b} are presented for all the
    panels with the time shift of $+0.4$\,s. 
\label{fig4}}
\end{figure*}

Upper panels of Fig.~\ref{fig3} plot the evolution of the
electromagnetic energy, $E_{\rm B}$, and the ratio of $E_{\rm B}$ to
the kinetic energy, $E_{\rm kin}$, as functions of time for the
systems of a black hole and a low-mass disk.  For all the models, the
electromagnetic energy increases exponentially with time in the early
stage of the evolution. The growth timescale, indicated by the black
dotted line in the left panels of Fig.~\ref{fig3}, agrees
approximately with the expression of Eq.~(\ref{growmax}): As this
equation indicates, for the models with the higher values of
$\sigma_{\rm c}$ and $\alpha_{\rm d}$, the growth timescale is
shorter.  It is also found that the curves of $E_{\rm B}$ oscillate
with time.  This is likely to be the reflection of the fact that there
are multiple oscillatory modes in the presence of the dynamo term;
during the change of polarity of the magnetic fields with multiple
modes, the field strength should change in time (see the discussion in
Sec.~\ref{sec2-2}).

After the early exponential increase of the electromagnetic energy,
the growth rate becomes smaller due to the back reaction of the matter
affected by the electromagnetic force.  Associated with this reaction,
the mass outflow sets in.  Eventually, the exponential growth is
stopped when the ratio of the electromagnetic energy to the kinetic
energy, $E_{\rm B}/E_{\rm kin}$, reaches $\approx 0.03$--$0.1$. At
this stage, the early-time mass ejection is most activated (see, e.g.,
Fig.~\ref{fig4}).  Subsequently, the ratio of $E_{\rm B}/E_{\rm kin}$ 
remains to be between $10^{-2}$ and $10^{-1}$ as often found in the
accretion disks with an equipartition state. This ratio is slightly
higher for the larger values of $\sigma_{\rm c}$ for this low-mass
disk model. In the stage of the relaxed value of $E_{\rm B}/E_{\rm
  kin}$, due to the angular momentum transport associated with the MHD
effect, a substantial fraction of the disk matter falls into the black
hole, and the disk mass decreases with time (see
Fig.~\ref{fig4}). Note that the disk mass also decreases partly (in
the present case $\sim 20\%$ of the initial disk mass) due to the mass
ejection. Associated with the mass infall and mass ejection, the
electromagnetic energy of the disk decreases with time, although the
ratio, $E_{\rm B}/E_{\rm kin}$, is preserved to be of $O(0.01)$
(cf.~Fig.~\ref{fig3}). The decrease rate of $E_{\rm B}$ is smaller for
smaller values of $\sigma_{\rm c}$ and $\alpha_{\rm d}$, because the
mass infall and ejection proceed more slowly.

We note that the upper two panels of Fig.~\ref{fig3} indicate that the
dependence of the electromagnetic-energy curve on the grid resolution
is weak.  Thus, the MHD evolution process of the disk is likely to be
captured well with the current grid resolutions.


The lower panels of Fig.~\ref{fig3} display the evolution of $E_{\rm
  B}$ and $E_{\rm B}/E_{\rm kin}$ for the high-mass disk models. It is
found that the evolution processes of these quantities are
qualitatively similar to those for the low-mass disk models.  The
dependence of the ratio of $E_{\rm B}/E_{\rm kin}$ on $\sigma_{\rm c}$
is weaker than that for the low-mass model, and it varies in a narrow
range approximately between 0.01 and 0.04.  For both low-mass and
high-mass disk models, the electromagnetic energy for the models with
$\alpha_{\rm d}=2\times 10^{-4}$ is smallest for $t \agt 1$\,s among
the models with the same disk mass. The reason for this is that the
mass ejection proceeds earlier for the larger value of $\alpha_{\rm
  d}$ (cf.~Figs.~\ref{fig4}(a) and \ref{fig6}(a)).  Thus the value of
$\alpha_{\rm d}$ controls the mass ejection timescale for the black
hole-disk system.

\subsubsection{Mass ejection}

Figure~\ref{fig4}(a) shows the evolution of the disk mass, the rest
mass that falls into the black hole, and ejecta mass.  For comparison,
numerical results derived by our viscous hydrodynamics
simulation~\cite{Fujiba20b} are presented together.  Since the mass
ejection is delayed by the growth timescale of the turbulent state in
the present MHD simulations, we shift the results for the viscous
hydrodynamics simulation by $+0.4$\,s in time. The evolution curves of
the mass outside the black hole indicate that, as in our viscous
hydrodynamics simulation, about 20\% of the initial disk mass is
likely to be ejected from the system and the rest of the matter falls
into the black hole for all the models. These fractions are in a good
agreement with those in viscous hydrodynamics. For models M10L75a and
M10L70, the mass ejection timescale is longer than the simulation
time, and thus, the ejecta mass does not settle to the final value in
$\sim 6$\,s. However the curves for the ejecta mass and disk mass are
similar to those in other MHD models, and hence, we may expect that
the ejecta mass will approach asymptotically $\sim 0.02M_\odot$.

The mass ejection timescale in the MHD simulations is shorter for the
larger values of $\sigma_{\rm c}$ and $\alpha_{\rm d}$ (see the curves
for models M10L80 and M10L75b in Fig.~\ref{fig4}(a)). This is natural
because for the larger values of these two parameters, the growth
timescale of the magnetic-field strength is shorter
(cf. Sec.~\ref{sec2-2}).  In addition, the $\alpha$-$\Omega$ dynamo
instability occurs for the shorter-wavelength modes with the larger
values of these parameters (cf.~Eq.~(\ref{fastD})), resulting in the
higher magnetic power and earlier mass ejection.  Although this
timescale depends on the choice of the parameters as in the case of
viscous hydrodynamics in which the mass ejection timescale depends on
the viscous coefficient, the timescale is universally seconds in the
reasonable choice of the parameters. It should be also mentioned that
the numerical results depend only weakly on the grid resolution,
although with the higher grid resolution, the mass ejection sets in
earlier perhaps due to the better-resolved magnetic-field growth (for
the dynamo instability as well as for the MRI).

Figure~\ref{fig4}(b)--(d) plot the evolution of the average values of
the specific entropy, electron fraction, and velocity for the
ejecta. Again, for comparison, we plot the results in our viscous
hydrodynamics~\cite{Fujiba20b} together with the time shift of
$+0.4$\,s. It is found that all the quantities take similar but
slightly different asymptotic values from those obtained in the
viscous hydrodynamics simulations irrespective of the values of
$\sigma_{\rm c}$ and $\alpha_{\rm d}$.  Slightly systematic
differences are found as follows: (i) The asymptotic value of the
average electron fraction of the ejecta in the MHD simulations is by
$\sim 0.05$ smaller than that in viscous hydrodynamics; (ii) The
average velocity of the ejecta in the MHD simulations is $\sim
0.1c$--$0.15c$, while it is $\sim 0.05c$ in viscous hydrodynamics (the
average velocity becomes high only for $\sigma_{\rm c}=10^8\,{\rm
  s}^{-1}$, but for others, it is universally $\approx 0.1c$).  The
second fact (ii) is in particular likely to be related to the
difference in the mechanisms of the mass ejection between MHD and
viscous hydrodynamics.  Thus, in the following, we discuss our
interpretation for this difference in the mass ejection mechanism in
more detail.

\begin{figure}[t]
  \includegraphics[width=84mm]{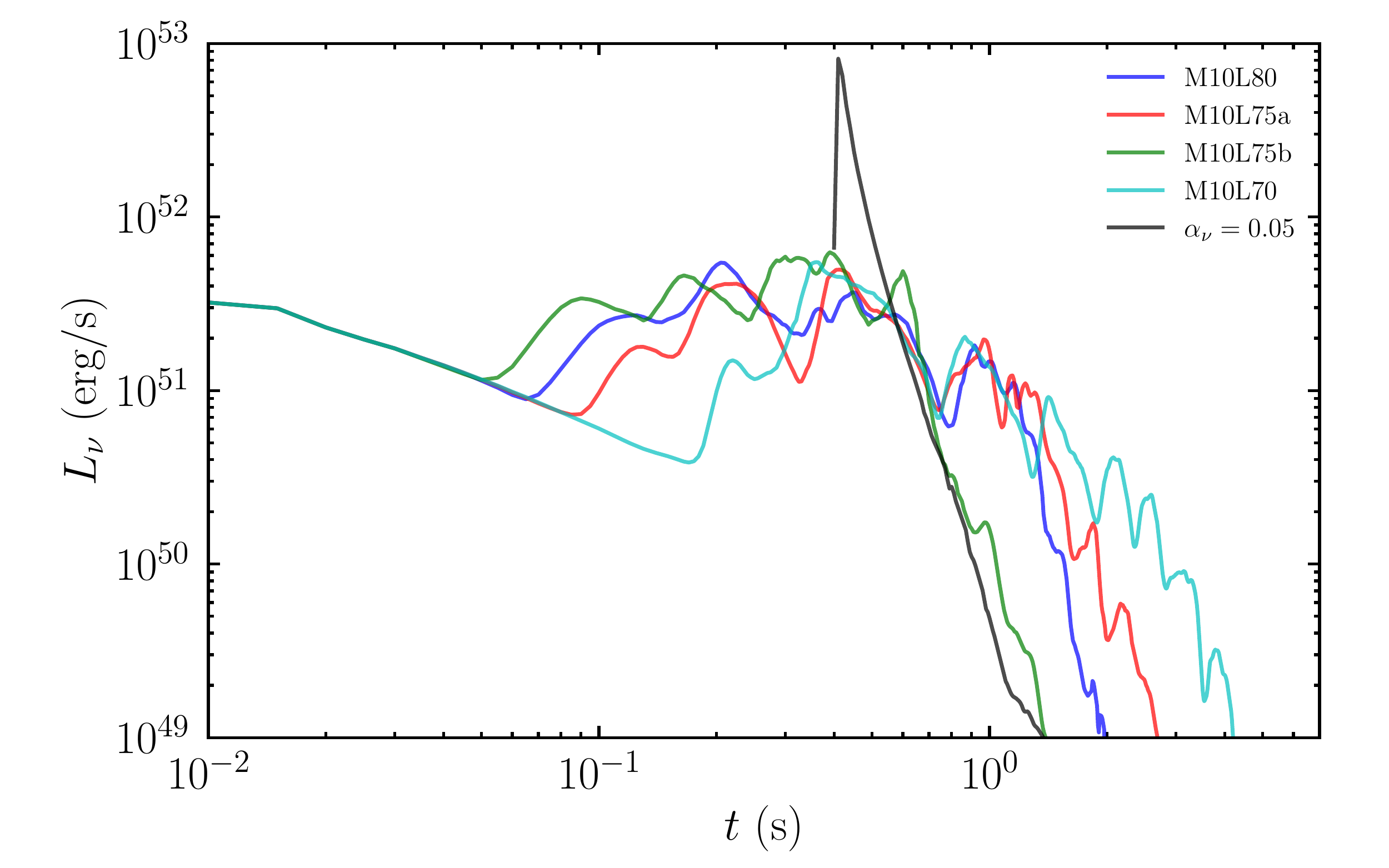}
  \includegraphics[width=84mm]{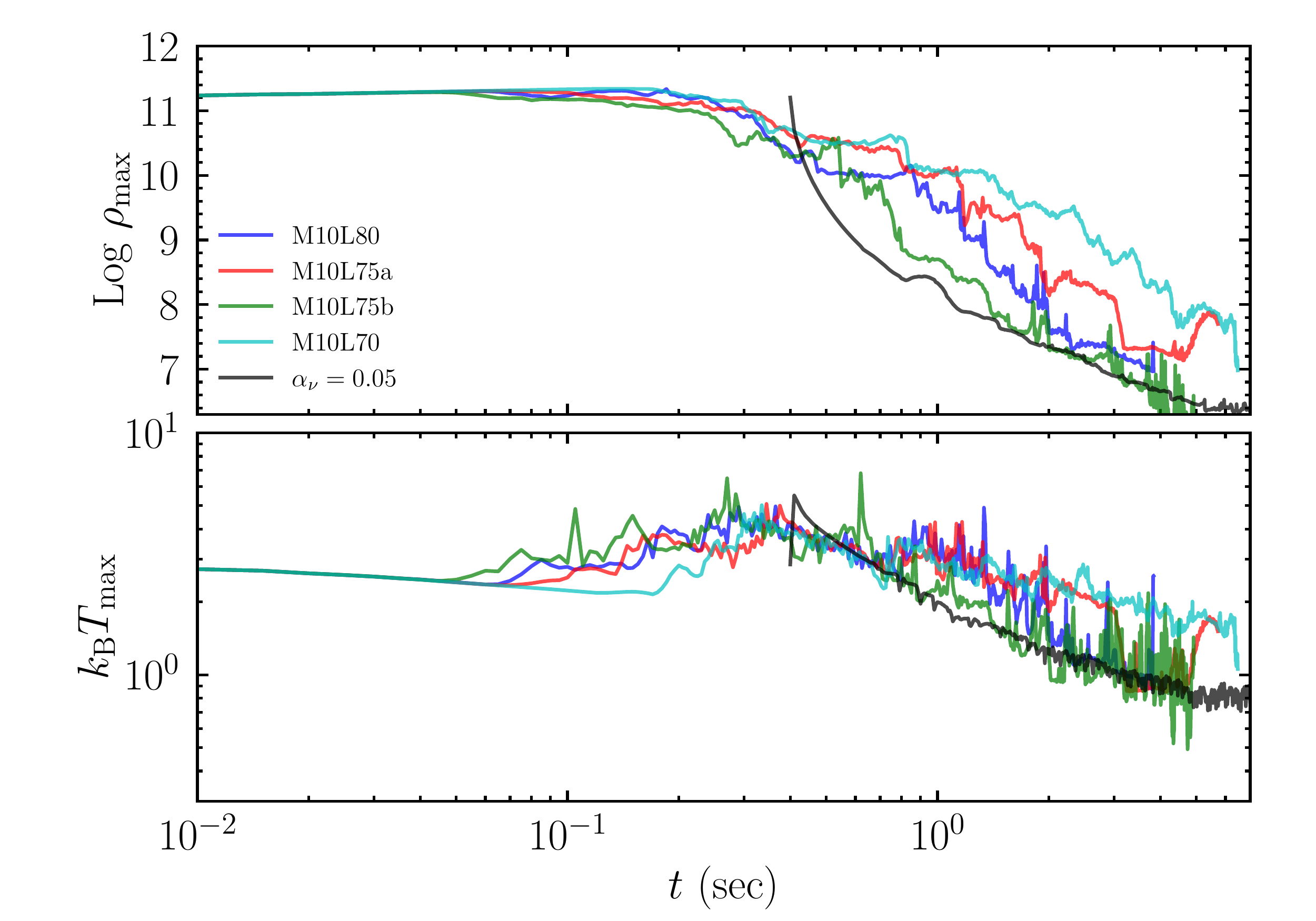}
\caption{Upper panel: Evolution of the neutrino luminosity for the
  low-mass disk model. Lower panel: Evolution of the maximum rest-mass
  density in units of ${\rm g/cm}^3$ and maximum temperature ($k_{\rm
    B} T_{\rm max}$) in units of MeV.  For comparison, the results by
  a viscous hydrodynamics simulation performed in
  Ref.~\cite{Fujiba20b} are presented with the time shift of
  $+0.4$\,s.
\label{fig5}}
\end{figure}

\begin{figure*}[t]
(a)\includegraphics[width=84mm]{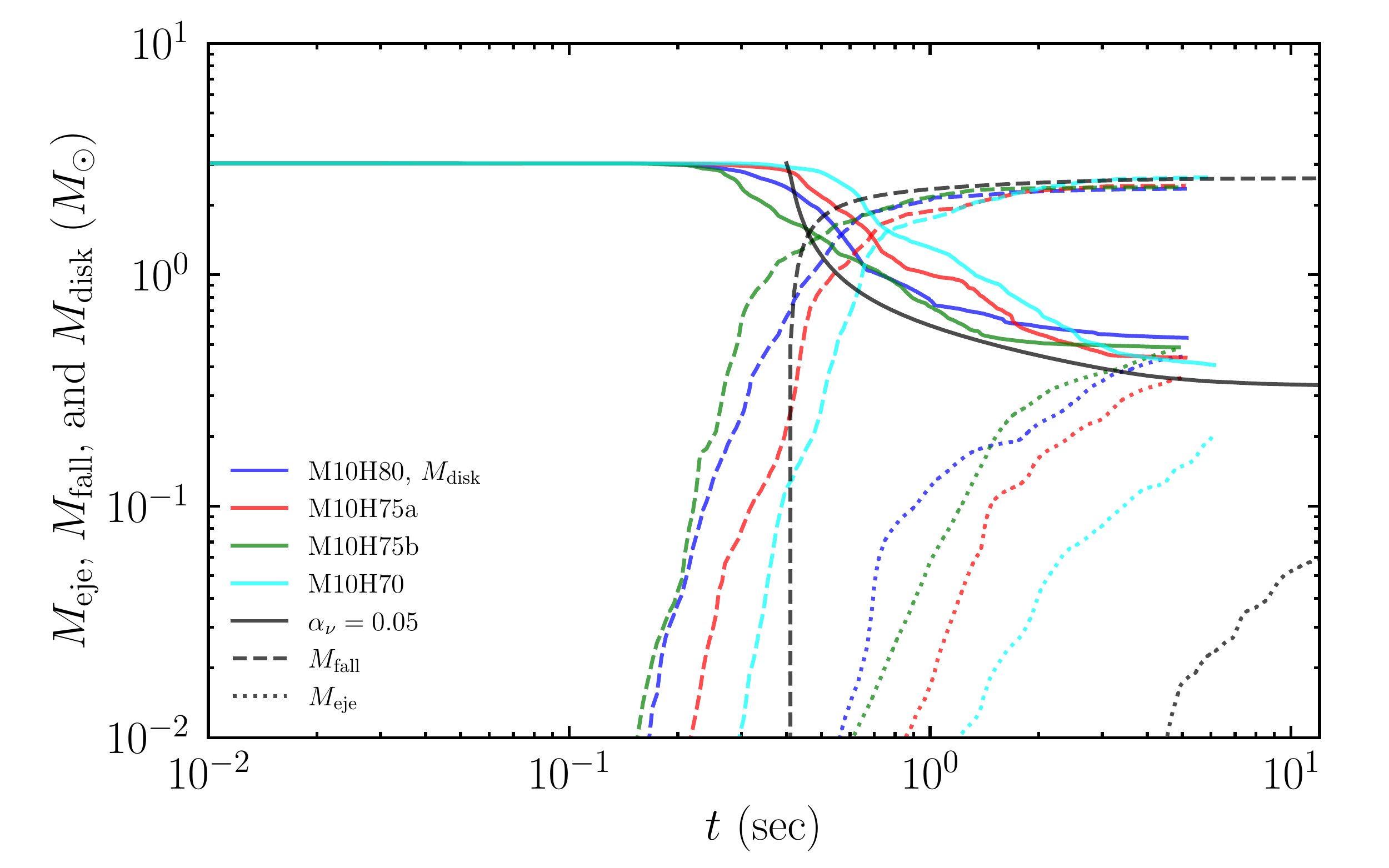}
(b)\includegraphics[width=84mm]{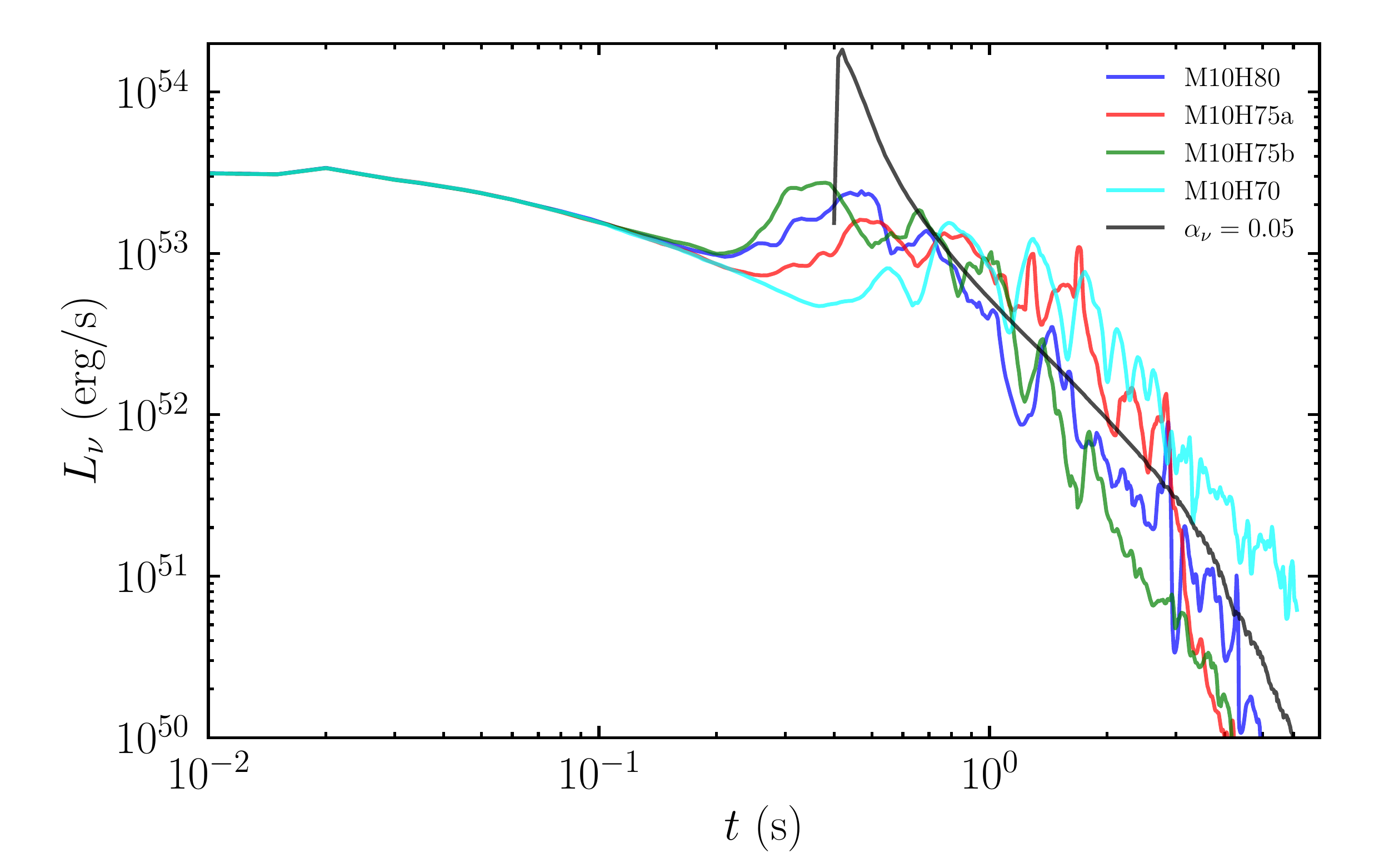}
(c)\includegraphics[width=84mm]{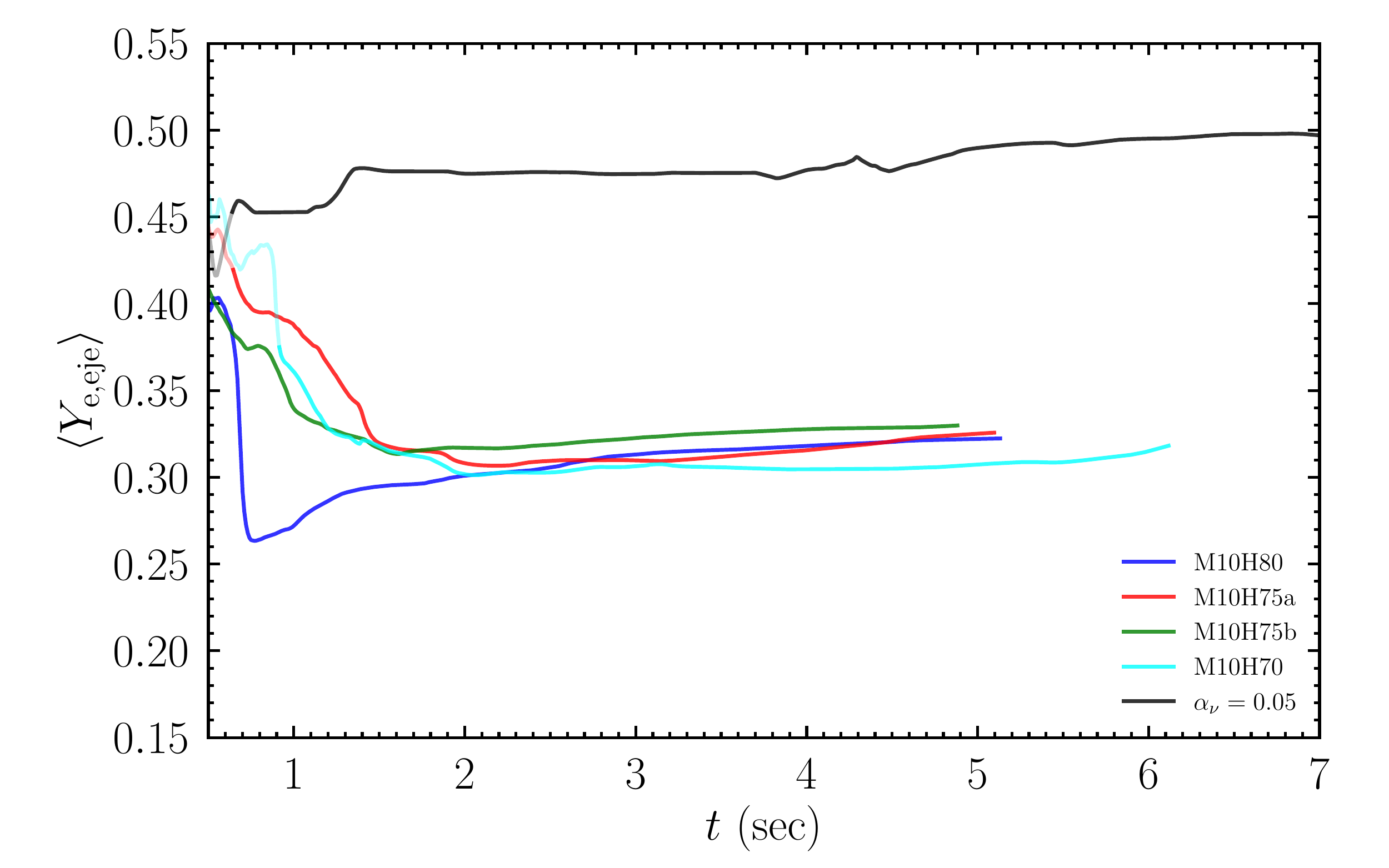}
(d)\includegraphics[width=84mm]{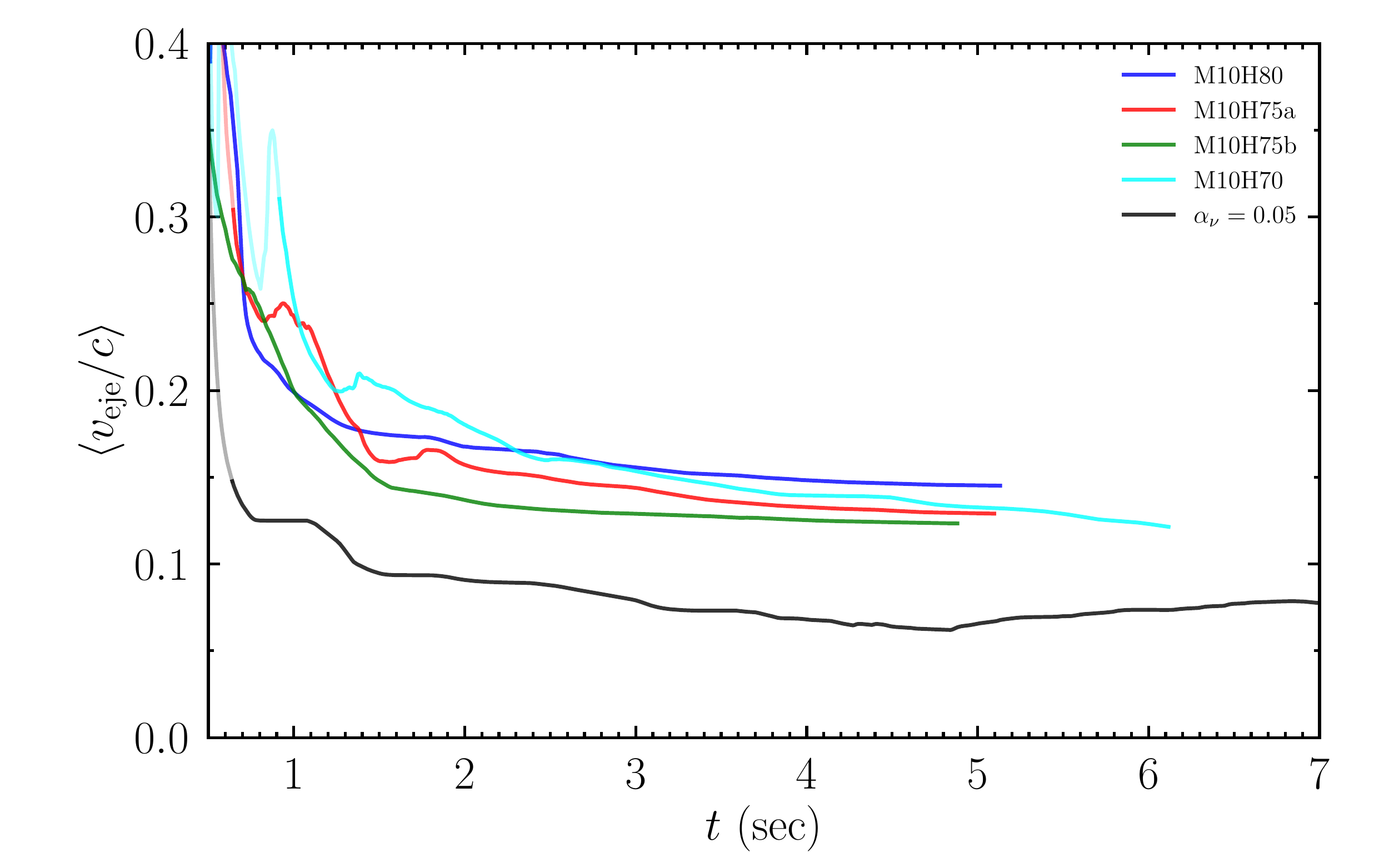}
  \caption{Evolution of several quantities for the high-mass disk
    surrounding a black hole. (a) Mass of the matter located outside
    the black hole (solid curve), that falls into the black hole
    (dashed curve), and ejected from the system (dashed curve); (b)
neutrino luminosity; 
    (c) the average value of
    $Y_e$ for the ejecta; (d) the average velocity of the ejecta.  For
    comparison, the results by a viscous hydrodynamics simulation
    performed in Ref.~\cite{Fujiba20b} are presented with a time shift
    of $+0.4$\,s for all the panels.
\label{fig6}}
\end{figure*}

For the case of viscous hydrodynamics, the mass ejection is driven by
the viscous heating after the disk is substantially evolved by the
viscous angular momentum transport by which the density and
temperature of the disk are decreased so significantly that the
neutrino cooling becomes inefficient~\cite{MF2013,Fujiba20}.  Here, it
is worth mentioning that in viscous hydrodynamics, the viscous heating
is the only channel for the mass ejection, and only when the thermal
energy gained by this heating is not significantly dissipated by some
other cooling processes such as the neutrino cooling, the mass
ejection can occur.  In MHD, on the other hand, the situation is
different: In this case, the turbulence induced by the MHD instability
enhances an effective viscosity and drives the angular momentum
transport and (effective) viscous heating in the same manner as in
viscous hydrodynamics, but there are additional magnetic-field effects
associated with the global motion of the magnetic-field lines. In
particular, in the presence of differential rotation, the
magneto-centrifugal~\cite{BP82} and magnetic-tower (e.g.,
Ref.~\cite{US85}) effects can play an important role in the mass
ejection. By these effects, the mass ejection can be driven even in
the absence of the effective viscous heating. In addition, the ejecta
as well as the matter in the outer region can be accelerated outward
by the magnetic power, even in the presence of an efficient cooling
process.


This interpretation is supported by observing the neutrino luminosity
as a function of time; see the upper panel of Fig.~\ref{fig5}.  In this
figure, we compare the neutrino luminosity in the MHD simulations with
the viscous hydrodynamics one.  Note that for the result of viscous
hydrodynamics (black solid curve), the time is shifted by $+0.4$\,s.
As shown in Ref.~\cite{Fujiba20}, in viscous hydrodynamics, the mass
ejection sets in when the neutrino cooling rate is much smaller than
the viscous heating rate; i.e., $L_\nu$ decreases below $\sim
10^{50}$\,erg/s.  On the other hand, in MHD, the mass ejection is
activated even in the stage with $L_\nu \sim 10^{51}$\,erg/s (see
Fig.~\ref{fig4}(a) and Fig.~\ref{fig5}). This is in particular the
case for models M10L75b and M10L80.  This early-time mass ejection is
likely to be driven primarily by the pure MHD effects.  For all the
models, however, the mass ejection is also active for the stage of low
neutrino luminosity. This late-time mass ejection is likely to stem
primarily from the effective viscous effect associated with the
enhanced turbulent viscosity and inefficient neutrino cooling. That
is, there are multiple channels for the mass ejection in
MHD~\cite{FTQFK19,Just2021}.

It is also found that the peak neutrino luminosity in MHD is at most
$10^{52}$\,erg/s while it is $\sim 10^{53}$\,erg/s in viscous
hydrodynamics. This indicates that the heating in MHD is not
enhanced as efficiently as in viscous hydrodynamics, as the neutrino
emission is closely related to the thermal energy of the matter. On
the other hand, the high-luminosity stage with $L_\nu \sim
10^{51}$\,erg/s continues for a long timescale of $\sim 1$\,s in MHD
while the stage with $L_\nu \agt 10^{51}$\,erg/s is only for $\sim
300$\,ms in viscous hydrodynamics. This indicates that the
instantaneous heating efficiency in MHD is not as high as that in
viscous hydrodynamics but the heating is continued constantly for a
long timescale. In other words, the timescale of the disk expansion
resulting from the (turbulent) viscous or MHD effect is not as short
as in viscous hydrodynamics. This fact is found from the evolution of
the maximum rest-mass density of the disk (see the bottom panel of
Fig.~\ref{fig5}): In the MHD simulations, the high value of the
maximum rest-mass density is preserved for a longer timescale than in
the viscous hydrodynamics simulation.  This is also reflected in the
fact that the maximum temperature in MHD is higher than that in
viscous hydrodynamics for $t \agt 1$\,s. We here note that for model
M10L80, the curves of $L_\nu$, $\rho_{\rm max}$, and $T_{\rm max}$ in
the late time are similar to those for the viscous hydrodynamics
model, but this is accidental: For model M10L80, the disk expansion
and mass ejection are driven mainly by the MHD effect with a short
timescale.

For the high-mass disk case, the difference in the mechanism for the
mass ejection between MHD and viscous hydrodynamics becomes even more
remarkable.  Figure~\ref{fig6}(a) shows the evolution of the disk
mass, the rest mass that falls into the black hole, and ejecta
mass. For comparison, again, numerical results derived by our viscous
hydrodynamics simulation with plausible viscous
parameters~\cite{Fujiba20b} are also presented. It is found that
irrespective of the values of $\sigma_{\rm c}$ and $\alpha_{\rm d}$
the asymptotic values of the mass for the matter located outside the
horizon in the MHD simulations are larger than that in the viscous
hydrodynamics one (and thus the mass that falls into the black hole in
the MHD simulations are smaller than that in viscous hydrodynamics).
In addition, the onset time of the mass ejection in the MHD simulation
is substantially earlier than that in the viscous hydrodynamics
one. As we already mentioned, in viscous hydrodynamics, the mass
ejection sets in only after the neutrino cooling becomes inefficient.
For the high-mass disk model, the neutrino luminosity is preserved to
be high for a long timescale of several seconds in both MHD and
viscous hydrodynamics (see Fig.~\ref{fig6}(b)). In viscous
hydrodynamics, this makes the onset time of the mass ejection later
than that in the low-mass disk model~\cite{Fujiba20b}. On the other
hand, in MHD, the difference in the onset time of the mass ejection is
not appreciable between the high-mass and low-mass disk models. This
indicates that the MHD effects, not viscous effects, play a primary
role in the mass ejection. In particular, for the high-mass disk case
for which the pressure of the disk is larger than that in the low-mass
disk case, the magnetic-field strength is enhanced to a higher level
in an equipartition stage (cf.~Fig.~\ref{fig3}), and furthermore, the
high-density region of the disk, which is located in its deep inside,
can play a role of an anchor for sustaining and swinging the field
lines of strong magnetic fields, leading to the increase of the mass
ejection efficiency via the magneto-centrifugal effect~\cite{BP82}.
This indicates that, in the presence of a high-mass dense object in
the central region, the efficiency of the mass ejection is enhanced.
This effect is in particular important in the presence of a neutron
star at the center (see Sec.~\ref{sec4}).

Figure~\ref{fig6}(c) and (d) plot the evolution of the average
electron fraction and velocity of the ejecta as in Fig.~\ref{fig4}.
In comparison with the viscous hydrodynamics results, a significant
difference is again found in the asymptotic values of the average
velocity: The average ejecta velocity in the MHD simulations is by a
factor of $\sim 2$ larger than the viscous hydrodynamics results in
the chosen ranges of $\sigma_{\rm c}$ and $\alpha_{\rm d}$ (in this
case the dependence of the ejecta velocity on $\sigma_{\rm c}$ is
weak). This is likely due to the enhanced MHD effects, in particular
to the magneto-centrifugal effect~\cite{BP82}, as mentioned above.  An
appreciable difference between the MHD and viscous hydrodynamics
results is also found in the average electron fraction. For the MHD
simulations, the average value of the electron fraction settles to
$\sim 0.35$ irrespective of the values of $\sigma_{\rm c}$ and
$\alpha_{\rm d}$, while in the viscous hydrodynamics simulation, it is
$\sim 0.5$. The reason for this high value in viscous hydrodynamics is
that for the high-mass disk model, the timescale of the mass ejection
from the disk is quite long, about several seconds, and during the
long-term evolution process of the disk, the density 
becomes low enough to decrease the electron degeneracy while
keeping relatively high temperature, and as a
result, the electron fraction is increased via the weak interaction
process~\cite{Fujiba20b}.  By contrast, in the MHD simulations, the
mass ejection is not primarily driven by the effective viscous process
resulting from the turbulent viscosity developed, but mainly by the
MHD activity such as magneto-centrifugal force, which significantly
shortens the mass ejection timescale (see Fig.~\ref{fig6}(a)). As a
result of these effects, the electron fraction of the ejecta remains
to be fairly low preserving the low values of the high-density state
of the disk.

\begin{figure}[t]
\includegraphics[width=84mm]{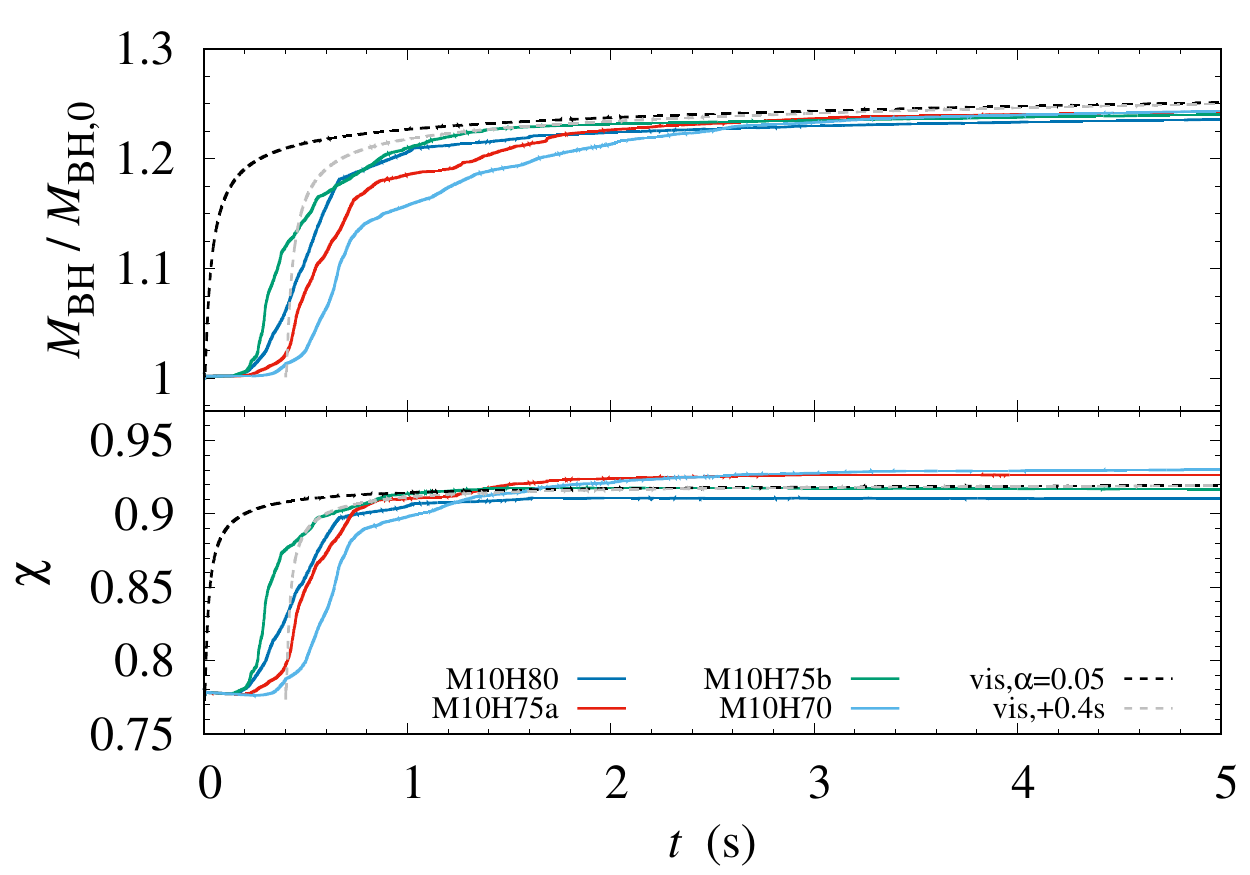}
\caption{The evolution of the mass and dimensionless spin of black
  holes for the high-mass disk model. The dashed curves are the
  results by a viscous hydrodynamics simulation performed in
  Ref.~\cite{Fujiba20b} (for the original result and for the results
  with the time shift of $+0.4$\,s).  In the upper panel, the
  black-hole mass in units of its initial mass is plotted.
\label{fig7}}
\end{figure}

\subsubsection{Evolution of black hole}

Although the mass ejection mechanism is different between the MHD and
viscous hydrodynamics models, the evolution process of the black hole
as a result of the mass accretion is similar in both approaches.
Figure~\ref{fig7} shows the evolution of the mass and dimensionless
spin of the black hole for all the high-mass disk models. For
comparison, we also plot the results obtained by the viscous
hydrodynamics simulation~\cite{Fujiba20b}. Irrespective of the MHD or
viscous hydrodynamics, the dimensionless spin increases with the mass
infall into the black hole and settles eventually to a saturated value
of $\approx 0.92\pm 0.01$ at $t \sim 3$\,s. Although the mass
accretion still continues for $t > 3$\,s and the black-hole mass
gradually increases, the dimensionless spin does not increase
significantly.  This result suggests the similarity of the angular
momentum transport by the MHD and viscous processes. The evolution
curves of the black-hole mass by the MHD simulations are also similar
to that by the viscous hydrodynamics simulation with a reasonable
viscous coefficient. The black-hole mass in the MHD simulations is
slightly smaller than that in the viscous hydrodynamics simulation.
The reason for this is that the fraction of the ejecta mass in the MHD
simulation is slightly larger than in the viscous hydrodynamics
simulations due to the MHD effects.

We note that for larger values of $\sigma_{\rm c}$ and $\alpha_{\rm
  d}$, the relaxed values of the dimensionless spin is slightly
smaller and the black-hole mass is slightly smaller.  This reflects
the fact that the outward angular-momentum transport effect and
resulting mass ejection are more efficient for the larger values of
$\sigma_{\rm c}$ and $\alpha_{\rm d}$.

\section{Evolution of a remnant of binary neutron star merger}\label{sec4}

\subsection{Setup}\label{sec4-1}

\begin{table}[t]
\caption{Initial conditions and setup for the numerical simulations of
  a binary neutron star merger remnant. For all the initial
  conditions, the total baryon mass is $M_*=2.95M_\odot$, the
  gravitational mass is $M=2.64M_\odot$, the total rotational kinetic
  energy is $E_{\rm kin}\approx 1.16 \times 10^{53}$\,erg, the
  electromagnetic energy is $E_{\rm B}\approx 2.61 \times
  10^{47}$\,erg, and the total angular momentum is $J=4.65 \times
  10^{49}$\,g\,cm$^2$/s.  }
\begin{tabular}{cccc} \hline
  ~~Model~~ & ~$\sigma_{\rm c}\,{\rm (s^{-1})}$~ & ~$\alpha_{\rm d}$~&
  ~~$\Delta x_0$\,(m)~~ \\
 \hline \hline
MNS80  & $1 \times 10^{8}$ & $1 \times 10^{-4}$ & 160, 200 \\
MNS75a & $3 \times 10^{7}$ & $1 \times 10^{-4}$ & 160, 200\\
MNS75b & $3 \times 10^{7}$ & $2 \times 10^{-4}$ & 160 \\
MNS75c & $3 \times 10^{7}$ & $5 \times 10^{-5}$ & 160 \\
MNS70a & $1 \times 10^{7}$ & $1 \times 10^{-4}$ & 160 \\
MNS70b & $1 \times 10^{7}$ & $2 \times 10^{-4}$ & 160 \\
 \hline
\end{tabular}
\label{table2}
\end{table}

We then turn our attention to the evolution of a binary-neutron-star
merger remnant, which is composed of a massive neutron star and a
torus.  As in our series of the
papers~\cite{Fujiba17,Fujiba18,Fujiba2020,SFS2021}, the initial
condition for the matter field is supplied from the result of a
simulation for the binary neutron star merger.  Specifically, we
employ the DD2-135 model of Ref.~\cite{Fujiba2020}: a merger remnant
of binary neutron stars with each neutron-star mass $1.35M_\odot$.
This model was already evolved in viscous hydrodynamics with the
(viscous) $\alpha$-parameter of 0.04 and the scale height of 10\,km in
a previous paper~\cite{Fujiba2020}.  We compare the results obtained
in the present GRRRMHD simulations with those in the viscous
hydrodynamics ones for several choice of $\sigma_{\rm c}$ and
$\alpha_{\rm d}$ in the following.

Again, we initially superimpose a purely toroidal magnetic field in a
high-density region of the remnant (both neutron star and torus) as
\beqn
\cB^T=\varpi \cB^\varphi
=A_0 \varpi z \,{\rm max}\left({P \over P_{\rm max}} - 0.01, 0\right). 
\label{initoro2}
\eeqn
The poloidal component of $\cB^i$ is set to be zero and the electric
field is determined by the ideal MHD condition of
$\cE^i=-\epsilon^{ijk}u_j\cB_k/w$.  The dependence on the coordinates,
$\varpi z$, in Eq.~(\ref{initoro2}) stems from the regularity
condition along the $z$-axis and the reflection anti-symmetry with
respect to the $z=0$ plane for $\cB^T$.  $A_0$ is a constant, and in
this work, we choose it so that the electromagnetic energy is $E_{\rm
  B} \approx 2.6\times 10^{47}$\,erg. We note again that the numerical
results depend very weakly on the initial field strength because the
electromagnetic field grows exponentially with time in the early stage
until a universal saturation level of the field strength is reached.
With the setting of Eq.~(\ref{initoro2}), the magnetic fields are
initially present in the massive neutron star and in the high-density
region of the torus.  As already mentioned in Sec.~\ref{sec3}, the
magnetic-field growth is driven purely by the dynamo instability for
the initial condition only with the toroidal magnetic field in the
axisymmetric simulation.

For the numerical simulation, the central region with $x \alt 30$\,km
and $z \alt 30$\,km is covered by the uniform grid of $\Delta
x_0=\Delta z_0=160$\,m or $200$\,m and outside this region, the grid
spacing is increased as $\Delta x_{i+1}=1.0075 \Delta x_i$ and $\Delta
z_{j+1}=1.0075 \Delta z_j$. We basically perform the simulations with
the higher grid resolution of $\Delta x_0=\Delta z_0=160$\,m, but to
confirm the weak dependence of the results on the grid resolution, for
selected models (MNS75a and MNS80), we also perform the simulations
with $\Delta x_0=\Delta z_0=200$\,m.  The models employed in this
paper are listed in Table~\ref{table2}. Unless otherwise stated, the
results with the higher-resolution setting are presented in the
following.  For all the models, the initial value of the kinetic
energy is $E_{\rm kin}\approx 1.16 \times 10^{53}$\,erg.

\subsection{Numerical results}\label{sec4-2}

\begin{figure*}[t]
\includegraphics[width=80mm]{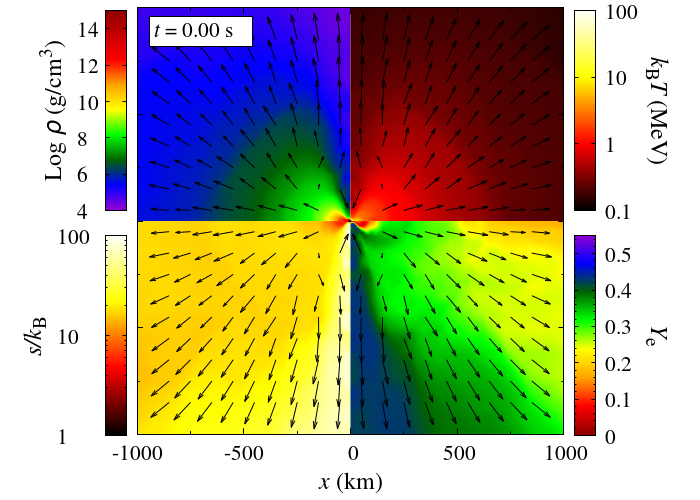}~
\includegraphics[width=80mm]{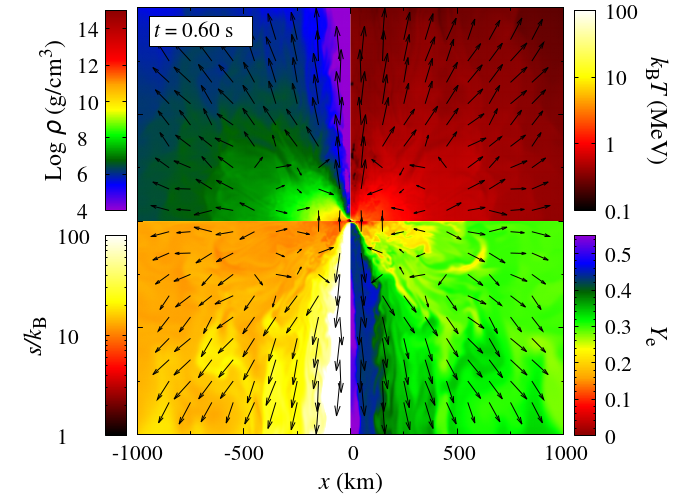} \\
\includegraphics[width=80mm]{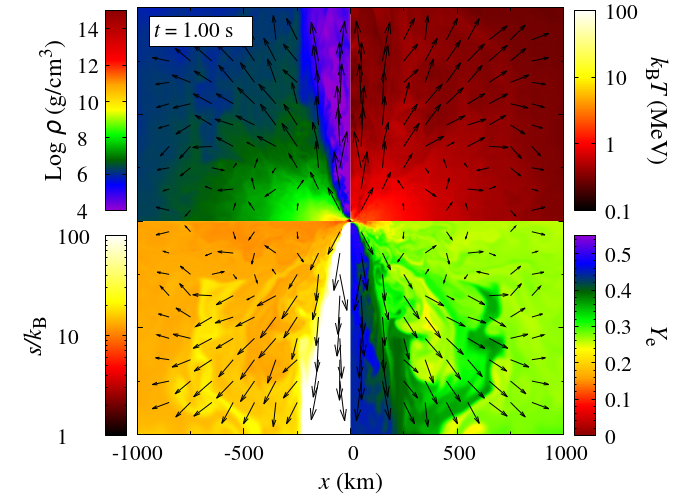}~
\includegraphics[width=80mm]{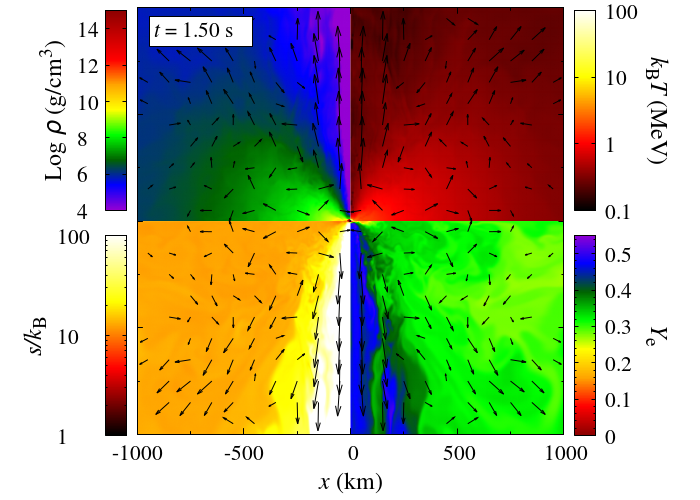} \\
\includegraphics[width=80mm]{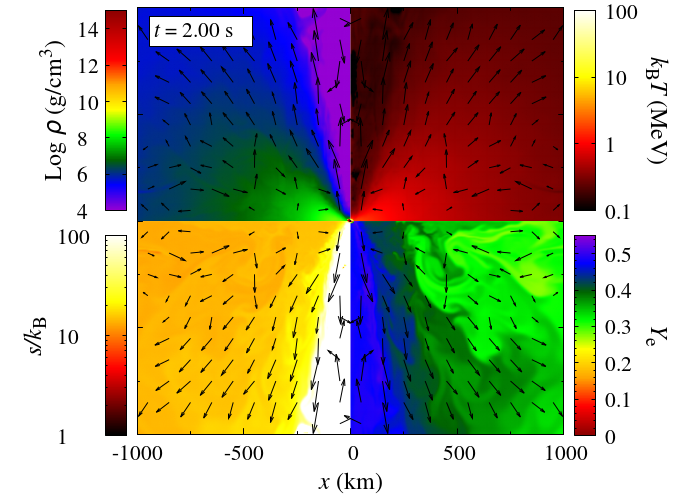}~
\includegraphics[width=80mm]{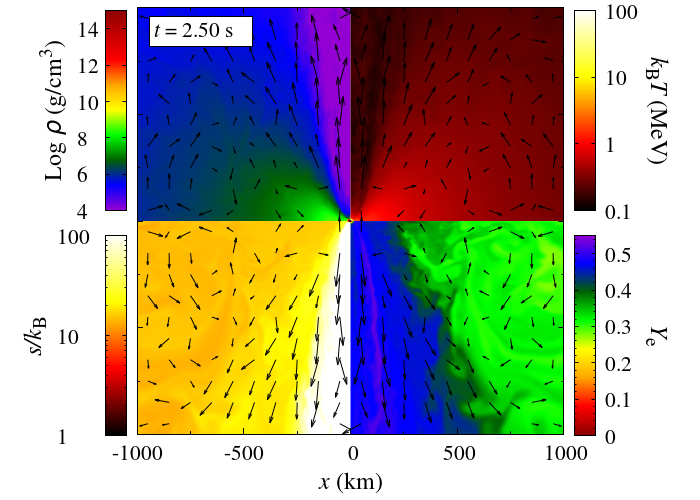} 
\caption{Snapshots of the rest-mass density in units of ${\rm
    g/cm}^3$, temperature ($k_{\rm B}T$ in units of MeV), specific
  entropy $s$ in units of $k_{\rm B}$, and electron fraction $Y_e$ at 
  selected time slices for model MNS75a with the high-resolution
  run. The arrows denote the velocity field of $(v^x, v^z)$.
\label{fig8}}
\end{figure*}
\begin{figure*}[t]
\includegraphics[width=84mm]{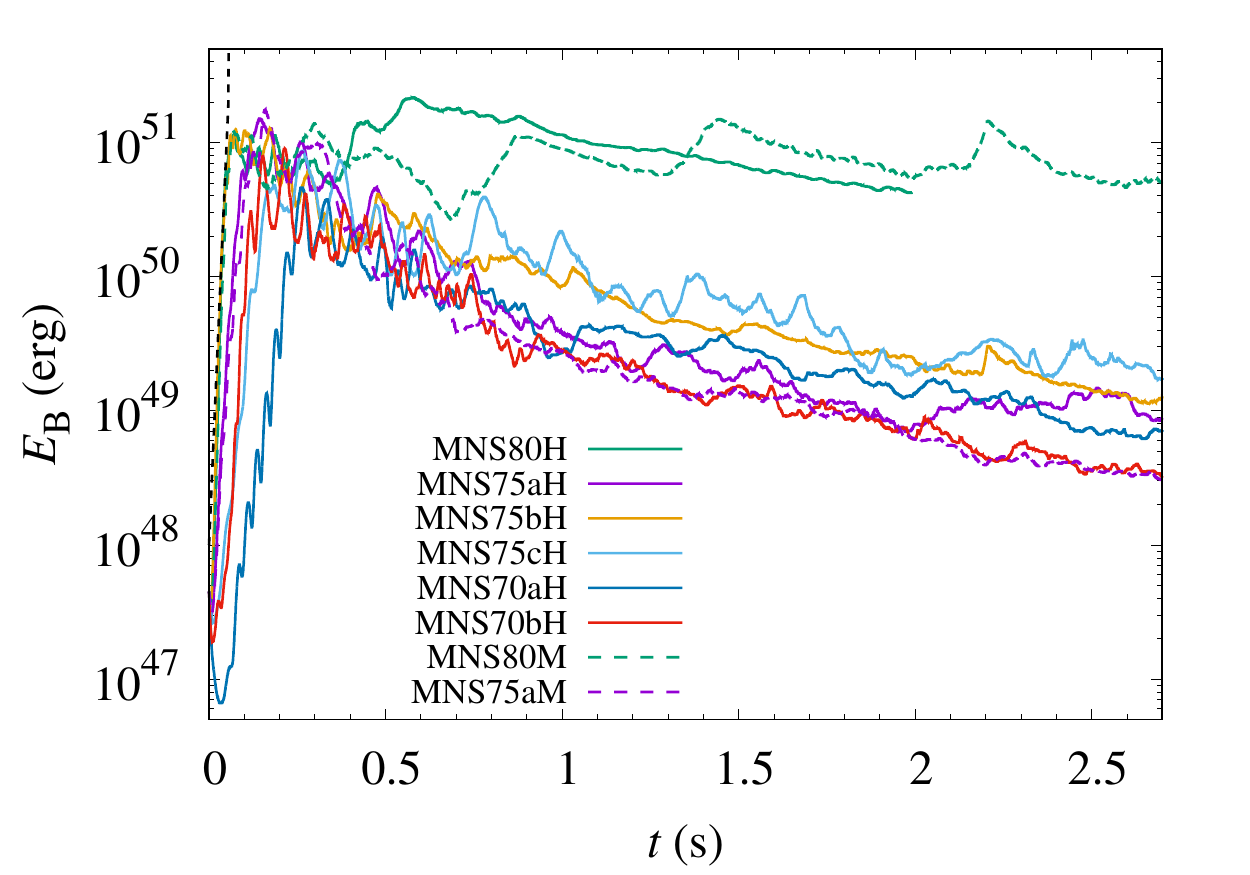}~~
\includegraphics[width=84mm]{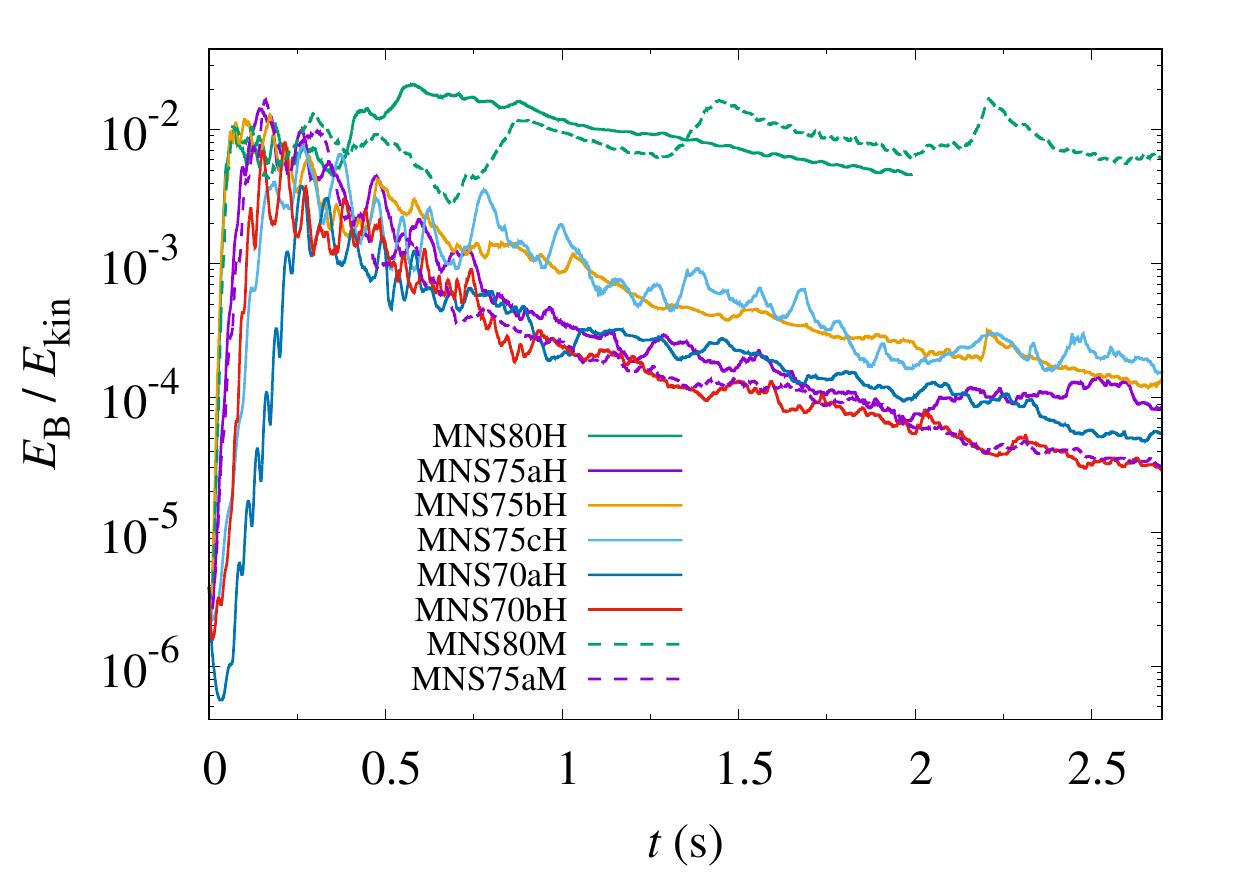}
\caption{Evolution of the electromagnetic energy (left) and ratio of
  the electromagnetic energy to the kinetic energy (right) for all the
  models listed in Table~\ref{table2}. The solid and dashed curves
  show the results with the high- and medium-resolution runs,
  respectively. The dotted black line in the left panel shows $\propto
  \exp(2\omega_{\rm max}t)$ with $\alpha_{\rm d}=10^{-4}$,
  $\sigma_{\rm c}=10^8\,{\rm s}^{-1}$, and $|S_\Omega|=10^3\,{\rm
    rad/s}$: cf.~Eq.~(\ref{growmax}).
\label{fig9}}
\end{figure*}
\begin{figure*}[t]
\includegraphics[width=170mm]{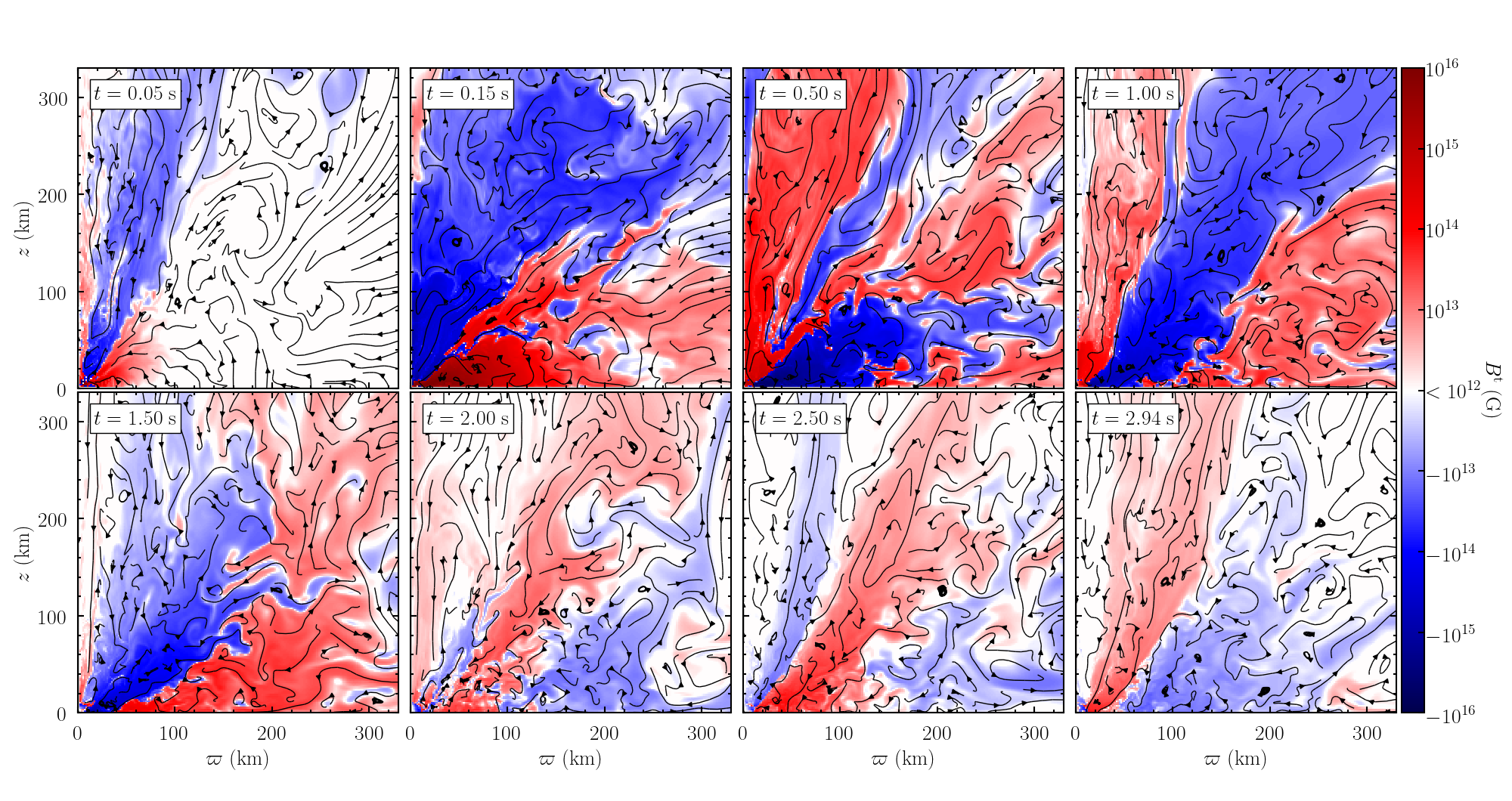}
\caption{The poloidal magnetic-field lines together with the toroidal
  magnetic-field strength (color profiles) at selected time slices for
  model MNS75a.
\label{fig10}}
\end{figure*}
\begin{figure*}[t]
\includegraphics[width=84mm]{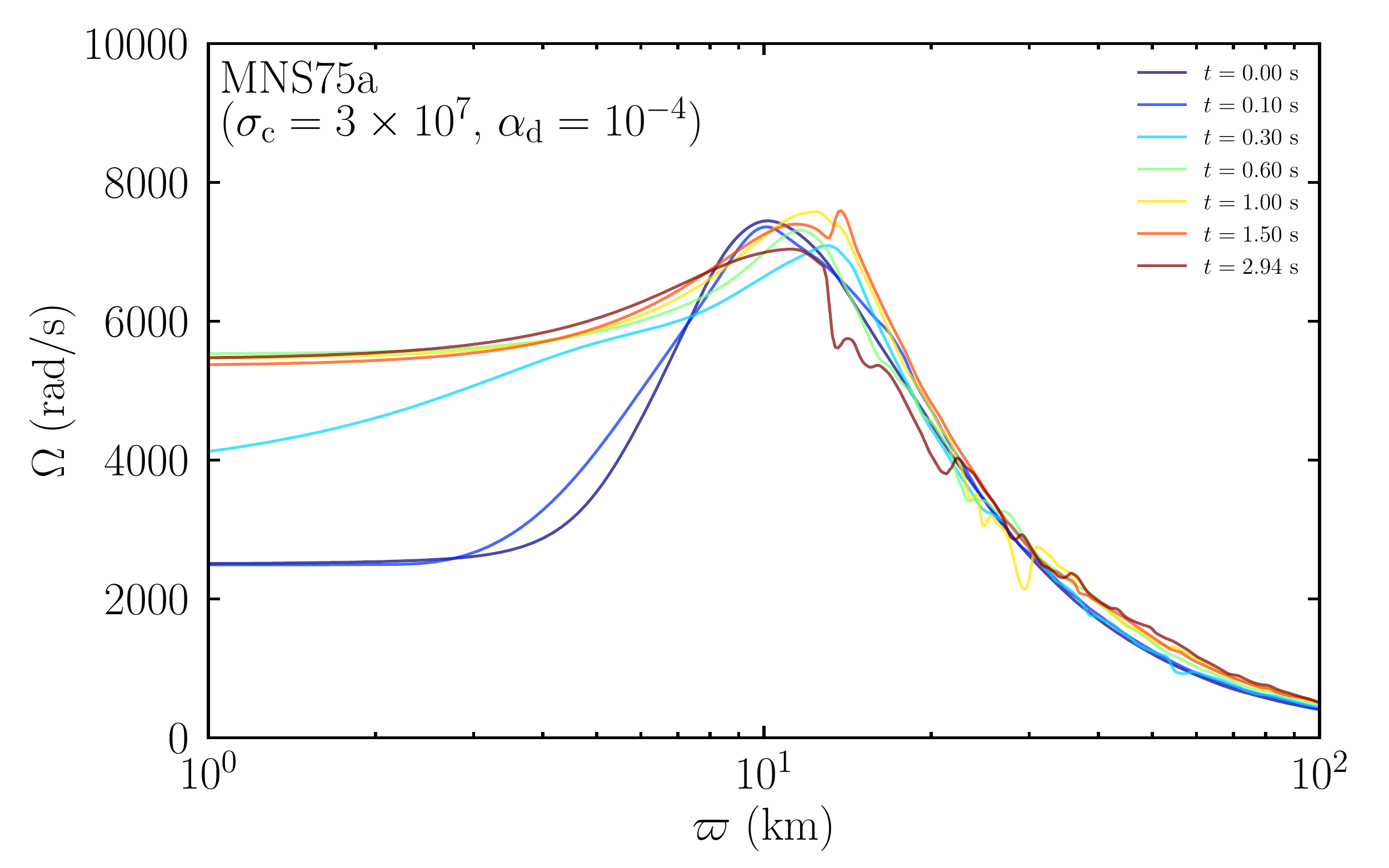}~~
\includegraphics[width=84mm]{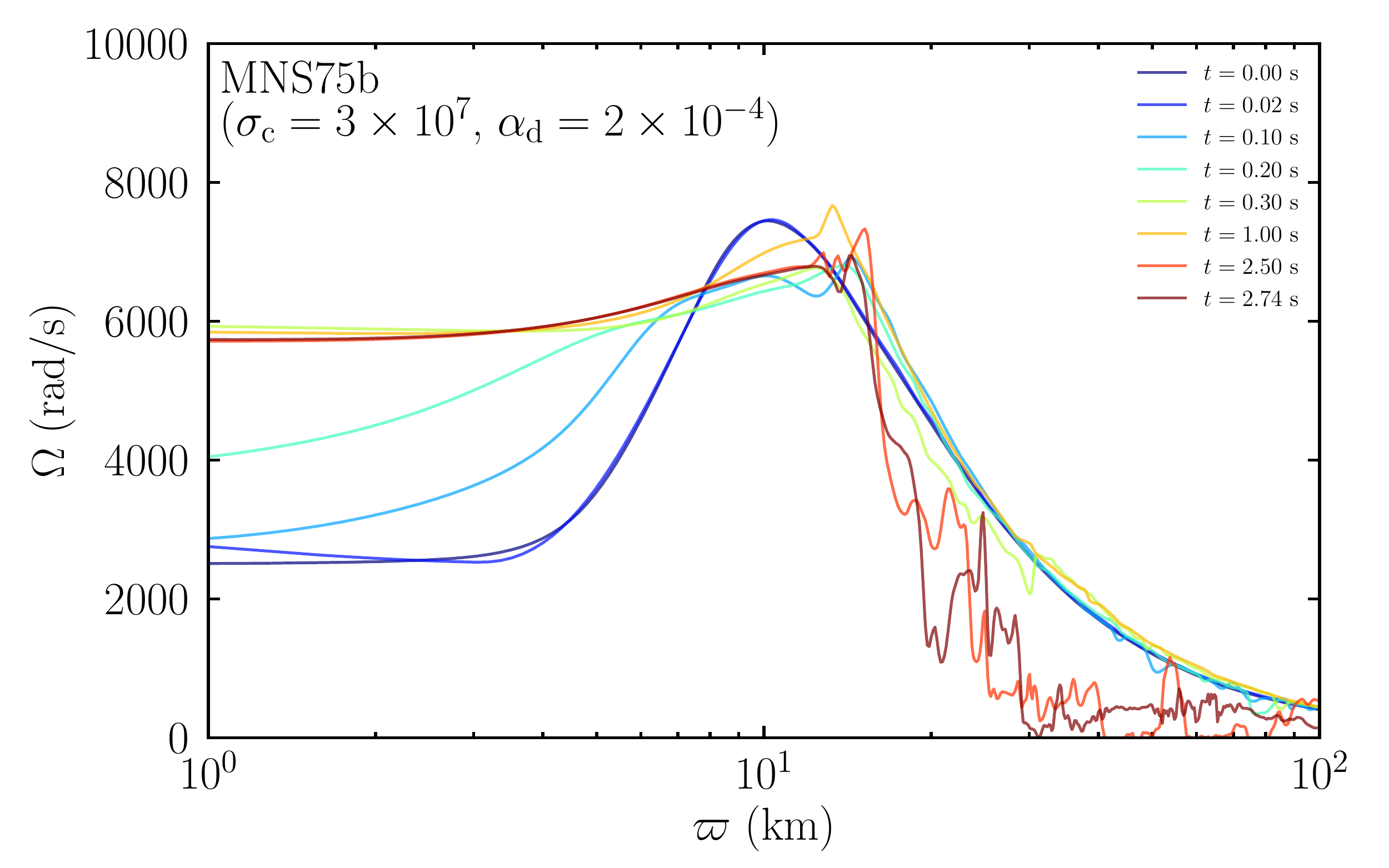}\\
\includegraphics[width=84mm]{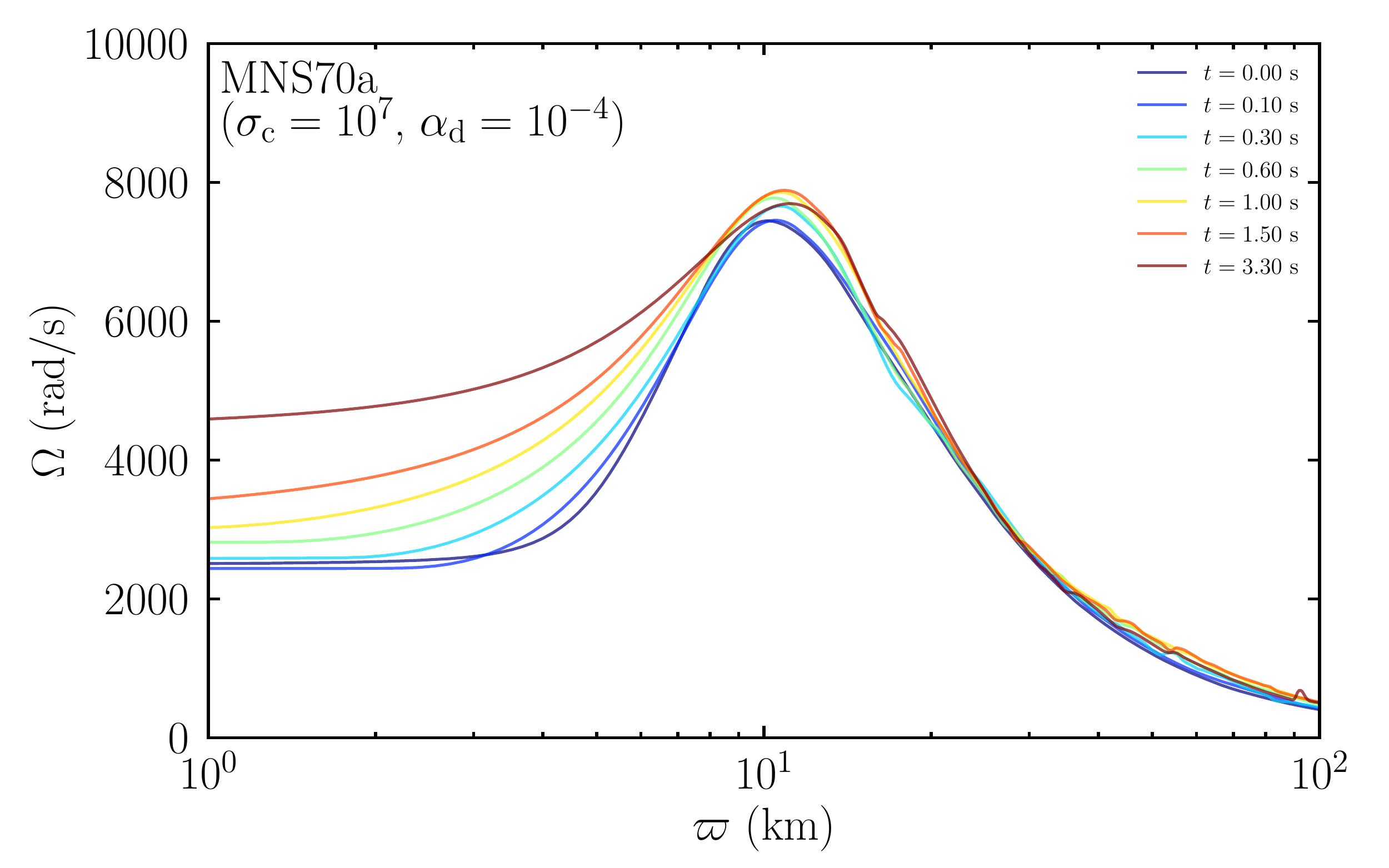}~~
\includegraphics[width=84mm]{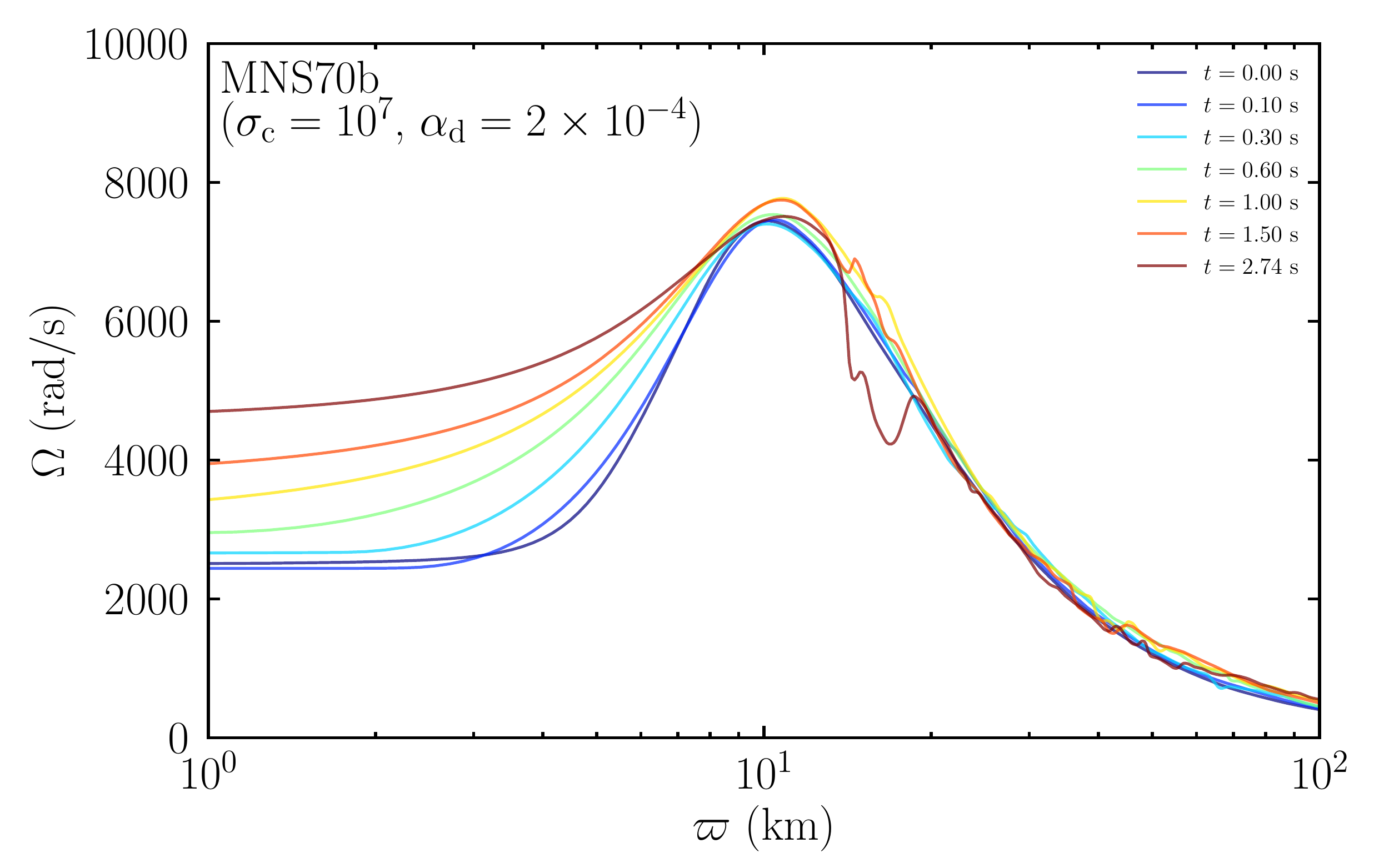}
\caption{Evolution of the angular velocity as a function of $\varpi$
  with $z=1$\,km for models MNS75a (top left), MNS75b (top right),
  MNS70a (bottom left), and MNS70b (bottom right). Note that for model
  MNS75b, the quasi-steady-state disk is absent for $t \agt 2$\,s because
  most of the disk matter is ejected by the magneto-centrifugal
  effect.
\label{fig11}}
\end{figure*}
\begin{figure*}[t]
(a)\includegraphics[width=84mm]{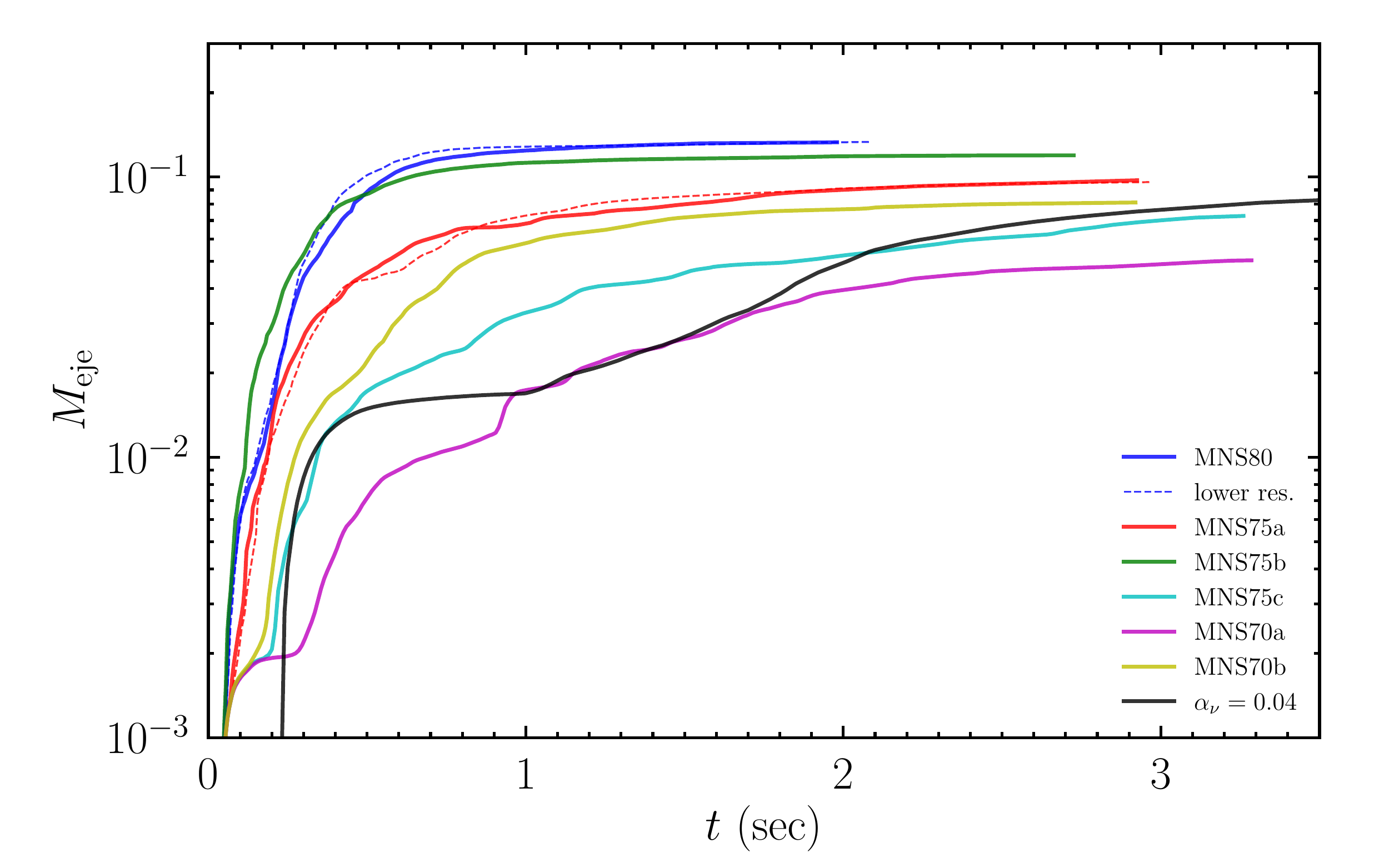}~~
(b)\includegraphics[width=84mm]{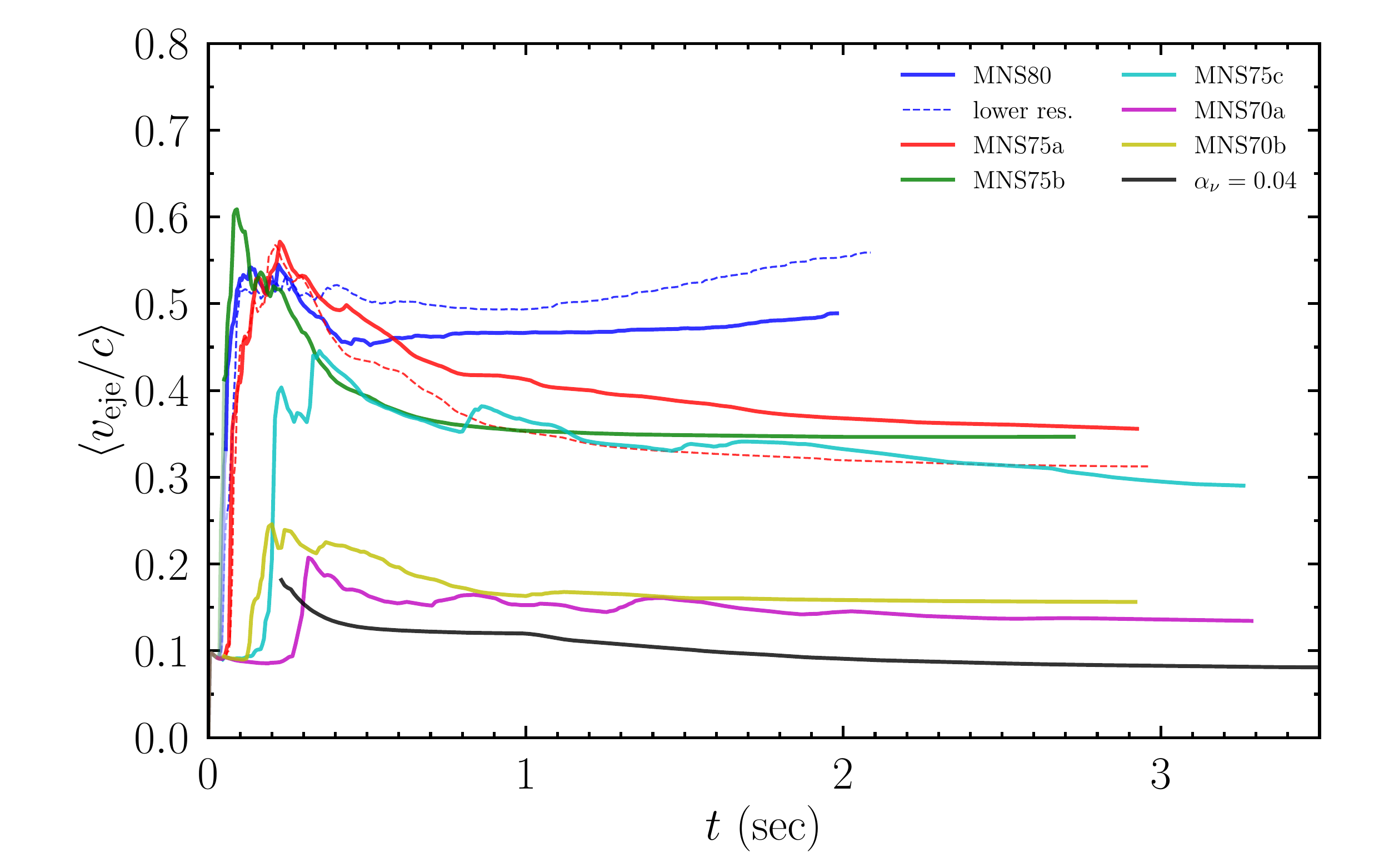} \\
(c)\includegraphics[width=84mm]{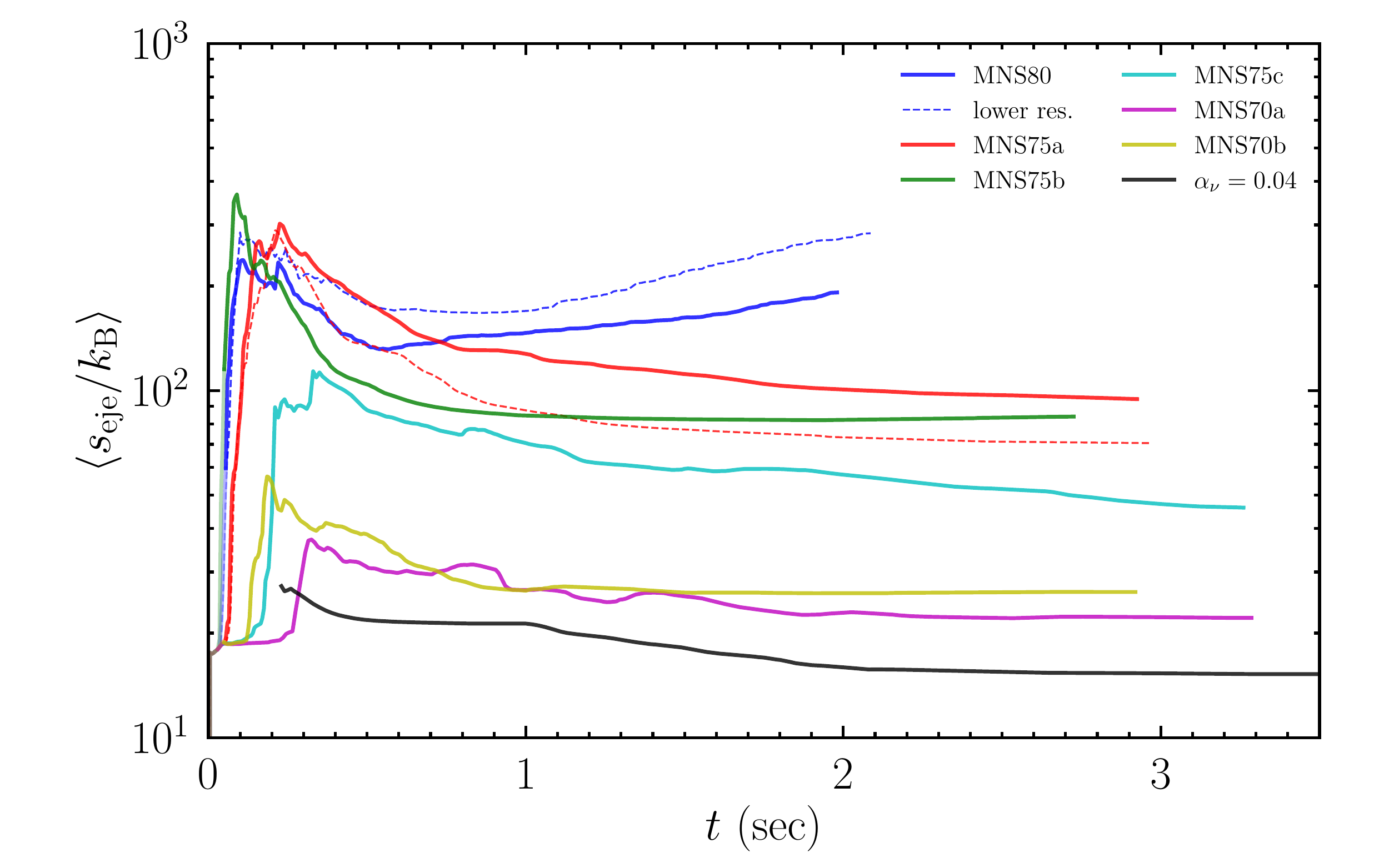}~~
(d)\includegraphics[width=84mm]{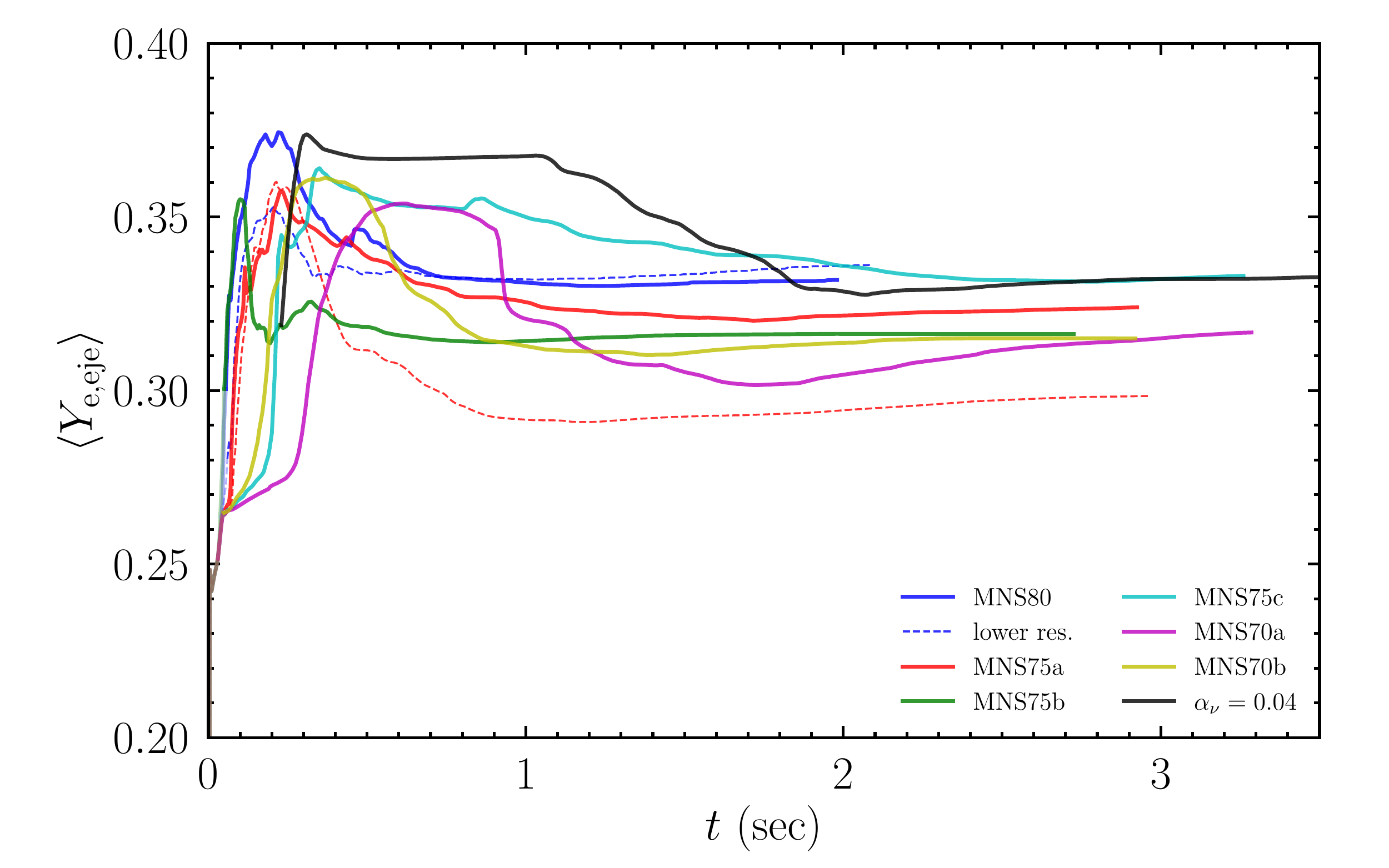}
  \caption{Evolution of (a) rest mass, (b) average velocity, (c)
    average specific entropy, and (d) average electron fraction of the
    ejecta for all the models listed in Table~\ref{table2}. The solid
    and dashed curves show the results with the high- and
    medium-resolution runs, respectively.  For comparison, the results
    for a viscous hydrodynamics simulation (model DD2-135M in
    Ref.~\cite{Fujiba2020}) are plotted together with the time shift
    of $+0.2$\,s.
\label{fig12}}
\end{figure*}

First of all, we display Fig.~\ref{fig8}, which shows the evolution of
the rest-mass density, the temperature, the specific entropy, and the
electron fraction for $\sigma_{\rm c}=3\times 10^7\,{\rm s}^{-1}$ and
$\alpha_{\rm d}=10^{-4}$ (model MNS75a). This illustrates a typical
evolution feature of the torus surrounding a neutron star in the
present MHD simulations: Due to the angular momentum transport and
heating associated with the MHD process caused by the enhanced
magnetic fields resulting from the dynamo action, the torus gradually
expands and the matter is ejected from the system spending $\sim
1$\,s. Eventually, the rest-mass density of the torus becomes much
lower than the initial value, and thus, the final outcome is a massive
neutron star with a low-density torus and its envelope. During the
evolution, it is also found that a funnel region with the half opening
angle of $10^\circ$--$15^\circ$ is established along the $z$-axis.  In
the funnel region, the electromagnetic pressure is comparable to or
larger than the gas pressure (i.e., with the plasma-$\beta$ of $\alt
1$). The resultant density profile is very similar to those in viscous
hydrodynamics (compare Fig.~\ref{fig8} with Fig.~1 of
Ref.~\cite{Fujiba2020}).  However, the mechanisms of the angular
momentum transport and heating in the torus and the mass ejection
process are quite different from those in viscous hydrodynamics. In
particular, the unique properties of the MHD processes can enhance the
efficiency of the mass ejection in the presence of strong magnetic
fields that have a base point (anchor) in the massive neutron star.
In the following, we pay particular attention to such unique
properties resulting from the MHD process.

Figure~\ref{fig9} shows the evolution of the electromagnetic energy
(left) and the ratio of the electromagnetic energy to the kinetic
energy ($E_{\rm B}/E_{\rm kin}$: right), respectively, for the entire
system of the remnant massive neutron star and the torus surrounding
it. As in the evolution for black-hole accretion disks, the
magnetic-field strength is initially amplified in the exponential
manner until $E_{\rm B}/E_{\rm kin}$ reaches $\sim 3\times
10^{-3}$--$1 \times 10^{-2}$, and then the amplification is saturated
irrespective of the values of $\sigma_{\rm c}$ and $\alpha_{\rm
  d}$. Here we note that the kinetic energy, $E_{\rm kin}$, is
dominated by that of the neutron star.  The typical maximum
magnetic-field strength is $\sim 10^{16}$\,G as found in
Ref.~\cite{SFS2021} (also comparable to that in the remnant of binary
neutron star mergers at a few ms after the merger~\cite{Kiuchi}).
Here the saturation occurs due to the fact that after the quick
magnetic-field amplification, the matter and magnetic flux start being
ejected from the neutron star and inner part of the torus primarily
toward the polar direction like in the magnetic-tower outflow (see
Fig.~\ref{fig10}; see also Ref.~\cite{SFS2021}), and the further
amplification is suppressed. These initial mass and magnetic-flux
ejections occur approximately when the maximum electromagnetic energy
is reached, i.e., at $\sim 0.2$--0.3\,s after the start of the
simulation.  This initial ejection is stronger for the larger values
of $\sigma_{\rm c}$ and $\alpha_{\rm d}$ because of the rapid growth
of the magnetic-field strength by the dynamo effect.  In particular,
for the cases of $\sigma_{\rm c}=10^8\,{\rm s}^{-1}$ and of
$\sigma_{\rm c}=3 \times 10^7\,{\rm s}^{-1}$ and $\alpha_{\rm
  d}=2\times 10^{-4}$ (models MNS80 and MNS75b, respectively), this
early mass ejection explosively occurs and becomes the dominant mass
ejection process among the entire evolution.

After the initial ejection toward the polar region, the magnetic
fields also spread to the equatorial region because of the strong
magnetic pressure and magneto-centrifugal effect.  Then, the matter
and magnetic-field flux start being outflowed toward a variety of the
directions and a global magnetic-field profile is established. After
the initial violent ejection, matter outflow quasi-steadily continues.
As we already mentioned, this early ejection is in particular
strong for models MNS80 and MNS75b. For these models, the torus 
around the massive neutron star is significantly disturbed during the
formation of the magnetic-field tower structure.

Because the poloidal magnetic field is developed by the
$\alpha$-$\Omega$ dynamo and associated outflow, the magnetic braking
is subsequently activated in the neutron star, and then, the degree of
the differential rotation in the neutron star becomes weak: inside it,
the angular velocity profile approaches a rigid state in particular
for models with $\sigma_{\rm c}\geq 3\times 10^7\,{\rm s}^{-1}$ (see
Fig.~\ref{fig11}). As a consequence, the value of $|S_\Omega|$ in the
inner region of the neutron star becomes small, leading to the
suppression of the $\alpha$-$\Omega$ dynamo there. However, the
differential rotation is still present in the outer part of the
neutron star and torus. Hence, the $\alpha$-$\Omega$ dynamo is still
active in the outer part of the system, preserves the turbulent state
of the torus, and induces the resulting mass ejection from the
torus. We note that the amplification of the magnetic field initially
occurs both in the neutron star and torus, but the total
electromagnetic energy is initially determined by that in the neutron
star. Also, the kinetic energy is always dominated by that of the
neutrons star and does not change significantly. For these reasons,
the shapes of the curves of $E_{\rm B}$ and $E_{\rm B}/E_{\rm kin}$
are similar to each other.

The evolution of the electromagnetic energy inside the neutron star
after the saturation of its growth depends strongly on the choice of
$\sigma_{\rm c}$, which determines the dissipation timescale (for
$S_\Omega=0$) given by 
\beqn
\tau_{\rm dis} &\approx&
    \left[{(kc)^2 \over 4\pi \sigma_{\rm c}}-\alpha_{\rm d} kc \right]^{-1}
\nonumber \\
&\approx & 0.1
\lambda_{20}
\alpha_{{\rm d},-4}^{-1}
\left(0.75\lambda_{20}^{-1}\alpha_{{\rm d},-4}^{-1}\sigma_{{\rm c},8}^{-1}-1
\right)^{-1}\,{\rm s}, ~~ \label{taudis}
\eeqn
where $\lambda:=2\pi/k$, $\lambda_{20}:=\lambda/(20\,{\rm km})$,
$\alpha_{{\rm d},-4}:=\alpha_{\rm d}/10^{-4}$, and $\sigma_{{\rm
    c},8}:=\sigma_{\rm c}/10^{8}\,{\rm s}^{-1}$, respectively. Note that if
$\tau_{\rm dis}$ is negative, the system is unstable for the
$\alpha$-dynamo with the corresponding wavelength.  Thus, for
$\sigma_{\rm c}=10^8\,{\rm s}^{-1}$ and $\alpha_{\rm d}=10^{-4}$, the
electromagnetic field in the neutron star can be preserved by the
unstable modes with $\lambda \agt 15$\,km, which is comparable to the
neutron-star radius, while for the smaller values of $\sigma_{\rm
  c}\leq 3\times 10^7\,{\rm s}^{-1}$ (and $\alpha_{\rm d} \leq 2
\times 10^{-4}$), the electromagnetic energy in the neutron star
should be dissipated in $\sim 0.1$\,s, because the modes with
$\lambda \alt 45$\,km decay.

Figure~\ref{fig9} indeed shows that for $\sigma_{\rm c}=10^8\,{\rm
  s}^{-1}$ (model MNS80), the electromagnetic energy is preserved
after the saturation of the field growth. On the other hand, for
$\sigma_{\rm c}\leq 3\times 10^7\,{\rm s}^{-1}$, the electromagnetic
energy decreases with time and its magnitude depends weakly on the
values of $\sigma_{\rm c}$ and $\alpha_{\rm d}$. The reason for this
is that in this later stage, the electromagnetic energy is dominated
by that of the torus in which the $\alpha$-$\Omega$ dynamo continues
to be active.  Indeed the magnitude of the electromagnetic energy,
$10^{48}$--$10^{49}\,{\rm erg}$, is in broad agreement with the black
hole-low-mass disk cases for which the order of the disk mass is the
same as that for the torus surrounding the neutron star (compare the
left panel of Fig.~\ref{fig9} with the top-left panel of
Fig.~\ref{fig3}).  Figure~\ref{fig10} also shows that (i) as in the
disk around the black hole (compare with Fig.~\ref{fig2}), the
polarity of the magnetic fields changes with time due to the dynamo
effect and (ii) due to the dissipation of the magnetic field in the
neutron star, the magnetic-field strength decreases along the $z$–axis
in the late stage of $t \agt 2$\,s. These results are essentially the
same as those in the black hole-disk case.  On the other hand, we do
not find clearly aligned structure for the magnetic-field lines in
the funnel region. Our interpretation for this is that the matter
outflow from the central region toward the polar region continuously
occurs and the magnetic-field structure is always disturbed in the
presence of the neutron star.

For $\sigma_{\rm c}=10^7\,{\rm s}^{-1}$ (models MNS70a and MNS70b),
the dissipation timescale of the magnetic fields in the neutron star
is so short that the magnetic braking effect is not very outstanding
as found in the bottom panels of Fig.~\ref{fig11}. That is, before the
sufficient growth of the poloidal fields in the neutron star
which activates the magnetic braking, the resistive effect becomes
important, and hence, the differential rotation is preserved for a
relatively long timescale. For these models, the effect of the neutron
star on the mass ejection becomes relatively minor (see below).


Associated with the magnetic-field amplification, the matter is
ejected from the merger remnant, primarily from the torus surrounding
the massive neutron star. However, this does not imply that the role
of the massive neutron star is not important for the mass ejection as
described in the following.

As already mentioned, by the enhancement of the magnetic-field
strength, the turbulent state is developed in the torus and the
enhanced magnetic-field force induces the mass ejection from the torus
(see, e.g., Figs.~\ref{fig8} and \ref{fig10}).  This situation is
qualitatively the same as that in the mass ejection from the black
hole-disk systems. However, in the presence of the neutron star in the
central region, the magneto-centrifugal effect~\cite{BP82} further
enhances the mass ejection efficiency because some of the
magnetic-field lines are anchored by the neutron star and the angular
velocity of the neutron star is higher than that of the torus. Because
of the presence of this additional effect, the total ejecta mass can
be higher than that in the viscous hydrodynamics simulation with
reasonable viscous parameters~\cite{Fujiba2020}, in particular for the
high values of $\sigma_{\rm c}$ (see Fig.~\ref{fig12}(a)) with which
the dissipation timescale of the magnetic fields in the neutron star
is longer.  In addition, the ejecta velocity is enhanced significantly
for $\sigma_{\rm c}\geq 3\times 10^7\,{\rm s}^{-1}$ (see
Fig.~\ref{fig12}(b)).  In particular for $\sigma_{\rm c}=10^8\,{\rm
  s}^{-1}$ (model MNS80), the average ejecta velocity becomes $\sim
0.5c$. Even for $\sigma_{\rm c}=3 \times 10^7\,{\rm s}^{-1}$, the
ejecta velocity is always by a factor of $\sim 2$ higher than that in
viscous hydrodynamics, and the kinetic energy of the ejecta $\approx
M_{\rm eje} v^2_{\rm eje}/2$ becomes $\agt 10^{52}$\,erg for models
MNS80, MNS75a, MNS75b, and MNS75c.  This implies that if the strong
magnetic-field lines anchored in the neutron star are present for a
few hundreds ms, the magneto-centrifugal force plays a significant
role in the mass ejection. By contrast, if the strong magnetic field
is present only for $\alt 100$\,ms (i.e., $\sigma_{\rm c}\leq
10^7\,{\rm s}^{-1}$), the magneto-centrifugal force is likely to be a
minor effect.  In particular, if the dynamo effect is not strong,
i.e., for MNS70a, the mass ejection efficiency in the MHD simulation
is weaker than in the viscous hydrodynamics case. However, if the
magnetic-field growth occurs in a short timescale, i.e., within $\sim
200$\,ms, the strong magnetic-field effect is universally observed
irrespective of the values of $\alpha_{\rm d}$ (compare the results
for models MNS75a, MNS75b, and MNS75c).

The ejecta of kinetic energy, $E_{\rm kin,eje}$, gained primarily by
the magneto-centrifugal force associated with the rotation of the
neutron star, should also obtain the angular momentum approximately by
$J_{\rm eje} \sim E_{\rm kin,eje}/\Omega$ where $\Omega$ is the
relaxed angular velocity of the neutron star which is $\Omega \approx
6000$\,rad/s (see Fig.~\ref{fig11}). With this process, the neutron
star should lose its angular momentum approximately by $J_{\rm
  eje}$. For $E_{\rm kin,eje}=10^{52}$\,erg, $J_{\rm eje} \sim 3
\times 10^{48}$\,erg\,s. The angular momentum of the neutron star is
approximately $3 \times 10^{49}$\,erg\,s.  Thus, $\agt 10\%$ of the
angular momentum of the neutron star is transported to the ejecta for
models with $\sigma_{\rm c}\geq 3 \times 10^7\,{\rm s}^{-1}$. This
result is reflected in the late-time decrease of the peak angular
velocity for models MNS75a and MNS75b (see the upper panels of
Fig.~\ref{fig11}).

With the higher value of $\alpha_{\rm d}$, the magnetic field is
amplified in a shorter timescale. As a result, the mass ejection sets
in earlier (compare the results of Fig.~\ref{fig12}(a) among models
MNS75a, MNS75b, and MNS75c and/or between models MNS70a and
MNS70b). However, the average velocity of the ejecta depends only
weakly on the value of $\alpha_{\rm d}$. This indicates that the
acceleration of the ejecta is induced primarily by the
magneto-centrifugal force related to the amplified magnetic-field
lines anchored in the neutron star.

\begin{figure}[t]
\includegraphics[width=86mm]{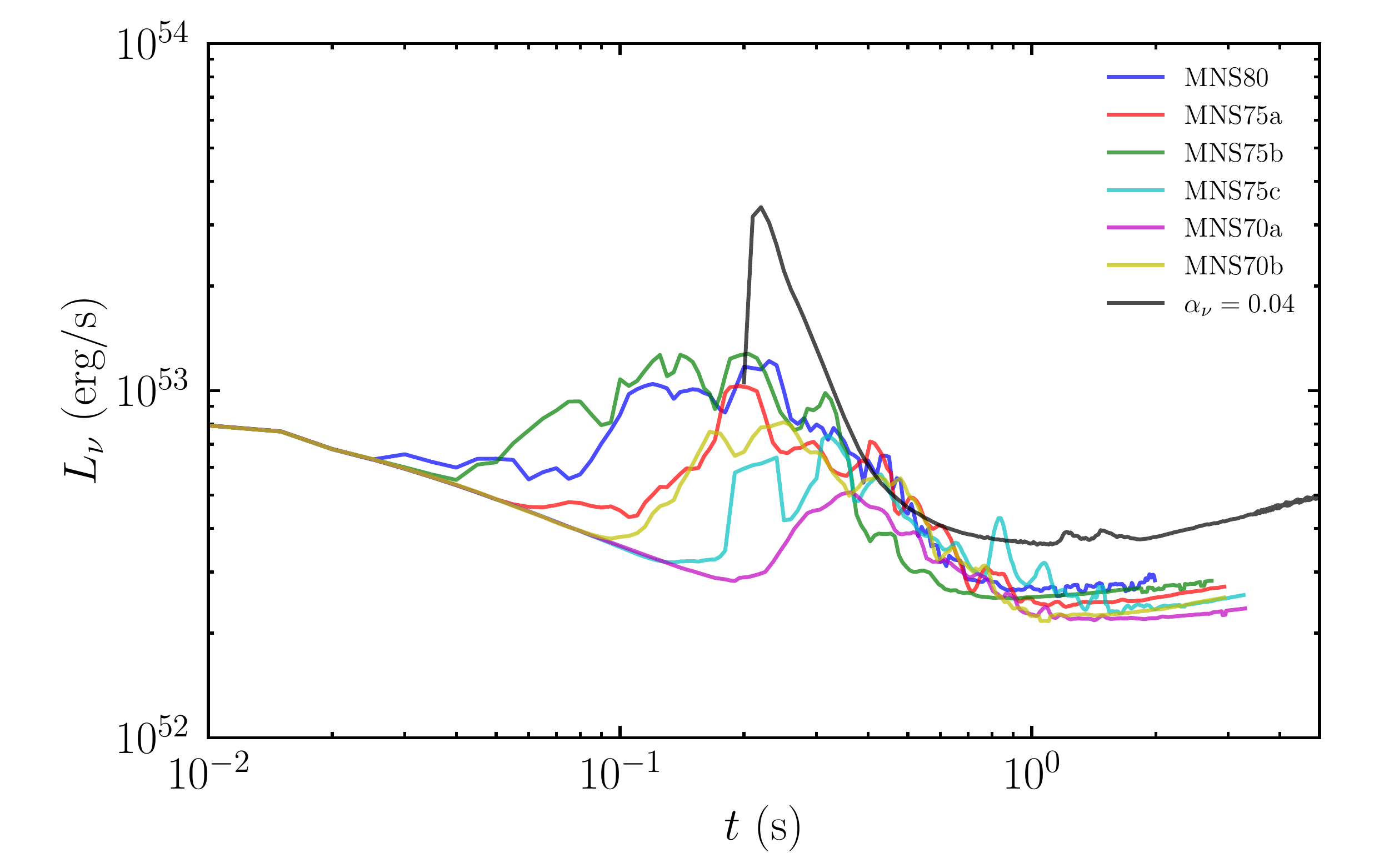}
\caption{Evolution of the neutrino luminosity for all the
  high-resolution models listed in Table~\ref{table2}.  For
  comparison, the result by a viscous hydrodynamics simulation
  performed in Ref.~\cite{Fujiba2020} is presented with the time
  shift of $+0.2$\,s.
\label{fig13}}
\end{figure}

Due to the violent magnetic-field activity and resulting shock
heating, the specific entropy of the ejecta is also significantly
enhanced, in particular for $\sigma_{\rm c} \geq 3\times 10^7\,{\rm
  s}^{-1}$ (see Fig.~\ref{fig12}(c)). This results from the efficient
shock heating by the MHD effect.  By contrast, the average electron
fraction is $\sim 0.3$, i.e., as high as that in the viscous
hydrodynamics simulation, irrespective of the values of $\sigma_{\rm
  c}$ and $\alpha_{\rm d}$ (see Fig.~\ref{fig12}(d)).  One reason for
this is that the majority of the matter is ejected from the outer part
of the torus for which the density is not so high that the electron
degeneracy is not very high, and thus, the neutron richness is only
moderately high.  The other reason is the presence of the strong
irradiation by neutrinos emitted from the massive neutron star.
Figure~\ref{fig13} plots the total neutrino luminosity as a function
of time. It is found that the neutrino luminosity in the MHD
simulations, $\sim 10^{53}\,{\rm erg/s}$, is only slightly smaller
than that in viscous hydrodynamics. Thus, the neutrino irradiation in
MHD can play a role as important as in viscous viscous hydrodynamics
for controlling the electron fraction of the matter surrounding the
neutron star and those ejected from it~\cite{Fujiba2020}.

Figure~\ref{fig13} also shows that for the larger values of
$\sigma_{\rm c}$ and $\alpha_{\rm d}$, the neutrino luminosity is
enhanced earlier. This indicates that the early growth of the
magnetic-field strength by the dynamo action contributes to enhancing
the shock-heating efficiency in the neutron star and resultant
neutrino emission efficiency. By contrast for $\sigma_{\rm
  c}=10^7\,{\rm s}^{-1}$ and $\alpha_{\rm d}=10^{-4}$ (model MNS70a),
the enhancement of the neutrino luminosity is minor, and the
neutron star appears to simply cool down. This
result is consistent with the interpretation that the MHD power is
reflected in the efficiency of the shock heating and neutrino
emission.

For $\sigma_{\rm c}=10^7\,{\rm s}^{-1}$ for which magnetic-field lines
anchored in the neutron star do not have a high field strength in the
main mass ejection stage, the results for the average velocity,
average entropy, and electron fraction are not very different from
those in viscous hydrodynamics irrespective of the value of
$\alpha_{\rm d}$, although the ejecta velocity is still higher than
that in viscous hydrodynamics.  In this case, the mass ejection is
driven primarily from the torus by the effective viscosity induced by
the MHD turbulence.  The weaker magnetic-field effect from the neutron
star is also reflected in relatively low neutrino luminosity as
mentioned above. Thus for models MNS70a and MNS70b, the properties of
the ejecta are similar to those in the viscous hydrodynamics
simulation~\cite{Fujiba2020}.  It is also worth mentioning that for
$\sigma_{\rm c}=10^7\,{\rm s}^{-1}$ and $\alpha_{\rm d}=2\times
10^{-4}$ (model MNS70b), the MHD and viscous hydrodynamics results are
similar to each other in the ejecta mass and the average value of the
electron fraction.

To summarize, the mass, velocity, specific entropy, and electron
fraction of the ejecta depend strongly on the strength of the global
magnetic fields anchored in the neutron star (and the lifetime of the
strong magnetic-field stage in the neutron star).  If the field
strength of the neutron star is preserved to be high enough for
several hundreds ms and the field lines are extended sufficiently far
from the remnant neutron star, the ejecta properties are affected
significantly by the magneto-centrifugal
effect~\cite{BP82}. Currently, the magnetic-field structure of the
merger remnant is not very clear.  Therefore one of the important
subjects in this research field is to clarify the magnetic-field
structure of the merger remnant by a long-term high-resolution MHD
simulation for neutron-star mergers.  Alternatively, if the future
electromagnetic observations for neutron-star mergers give us
information for the ejecta velocity, we may be able to learn the MHD
activity in the merger remnant.

\section{Summary}\label{sec5}

We performed GRRRMHD simulations incorporating a mean-field dynamo
term for black hole-disk systems and for a merger remnant of binary
neutron stars composed of a massive neutron star and a torus, paying
particular attention to the $\alpha$-$\Omega$ dynamo effect.  We
compared the new results with those previously obtained in our viscous
hydrodynamics simulations~\cite{Fujiba2020,Fujiba20,Fujiba20b} and
clarified the specific MHD effects.  For the system of a black hole
and a low-mass disk, it is found that the results of the MHD
simulations agree broadly with those of viscous hydrodynamics
simulations: The mass of the ejecta as well as the total mass that
falls into the black hole by the two approaches agree approximately
with each other. One clear difference is found in the average velocity
of the ejecta.  For the viscous hydrodynamics case, the average
velocity of the ejecta is smaller than $0.1c$ irrespective of the
viscous coefficient employed widely in our previous
studies~\cite{Fujiba20,Fujiba20b}.  By contrast, the average velocity
of the ejecta in the present MHD simulations becomes 0.10--$0.15c$.
Furthermore, mass ejection in the MHD simulations occurs relatively
earlier than that in viscous hydrodynamics, because not only the
effective viscosity effect resulting from the MHD turbulence but also
the magnetic force associated with the global magnetic fields
developed by the dynamo action plays an important role for the mass
ejection.  Associated with the enhancement of the ejecta velocity and
earlier mass ejection, the average value of the electron fraction is
slightly decreased.  The reason for this is that the ejecta is
generated relatively in early time from the disk, i.e., before the
matter significantly experiences the weak interaction effects, and is
also less subject to the irradiation of neutrinos emitted from the
accretion disk.  All these results are consistent with those found in
the previous MHD
simulation~\cite{SM17,FTQFK19,Miller19,Just2021}, although in
our results the velocity is not still extremely high and the average
value of the electron fraction is only mildly neutron rich as $\langle
Y_e \rangle \sim 0.3$.

For the case of a black hole and a high-mass disk, the disagreement
between the results of the MHD and viscous hydrodynamics simulations
is more remarkable. For this case, the mass ejection in MHD occurs
much earlier than in viscous hydrodynamics and the average velocity of
the ejecta in MHD is also appreciably larger than that in viscous
hydrodynamics. In addition, the average value of the electron fraction
is $\sim 0.35$ in MHD while it is $\sim 0.5$ in viscous
hydrodynamics. The main reason for this difference is that the
magneto-centrifugal effect plays a more significant (perhaps primary)
role in the mass ejection and ejecta acceleration than in the low-mass
disk case. In the presence of a massive disk, it is likely that the
magnetic field lines are anchored in the dense region of the torus,
and thus, the swinging of the global field lines and resultant ejecta
acceleration become more efficient.

For the massive neutron star-torus case, we also find a significant
enhancement in the mass and average velocity of the ejecta due to the
MHD effect, in particular for higher values of $\sigma_{\rm c} \geq 3
\times 10^7\,{\rm s}^{-1}$, i.e., for the cases that strong and global
magnetic-fields are preserved for several hundreds ms, in comparison
with that in the viscous hydrodynamics simulation.  For the high
values of $\sigma_{\rm c}$, strong magnetic-field lines anchored in
the neutron star, which is rotating more rapidly than the surrounding
matter such as torus, are preserved for several hundreds ms, and
therefore, by the magneto-centrifugal force, which is absent in
viscous hydrodynamics, the mass ejection is enhanced and also ejecta
are accelerated.  The present numerical results indicate that the
kinetic energy of the ejecta exceeds $10^{52}$\,erg/s for models
with $\sigma_{\rm c} \geq 3 \times 10^7\,{\rm s}^{-1}$. By contrast,
for $\sigma_{\rm c}=10^7\,{\rm s}^{-1}$, the magnetic field in the
neutron star is dissipated by the resistivity in a short timescale of
$\alt 100$\,ms, and hence, the magneto-centrifugal effect does not
become as significant as for $\sigma_{\rm c} \agt 3\times 10^7\,{\rm
  s}^{-1}$. For this high-resistivity case, the mass ejection proceeds
primarily through the effective viscous process resulting from the MHD
turbulence in the torus and the properties of the ejecta are similar
to those in viscous hydrodynamics.  This suggests that the
magnetic-field strength of the neutron star and global structure of
the magnetic field lines can primarily determine the ejecta properties
such as their mass and velocity. To clarify this point, we need a
self-consistent long-term simulation from the merger throughout the
post-merger evolution with the duration of seconds in the future.


The primary message of this paper is that if the strong amplification
of the magnetic fields occurs inside the remnant neutron star and a
global magnetic-field structure is established outside it, the ejecta
velocity can be much higher than that in the absence of the magnetic
fields~\cite{Fujiba2020}.  In our recent paper~\cite{Kawaguchi2021},
we derived light-curve models of kilonovae from the
binary-neutron-star merger remnant composed of a long-lived massive
neutron star and a torus. This work was based on the results of
viscous hydrodynamics of Ref.~\cite{Fujiba2020}. If we take into
account the MHD effects in the post-merger evolution, the light curve
of the kilonovae is likely to be modified significantly. Specifically,
the kilonovae, by high-velocity ejecta, are likely to shine earlier
and become bluer.  Also, the peak luminosity will be larger.  We plan
to explore the kilonova light curves using the numerical models
obtained in this work. The synchrotron emission generated during
sweeping the interstellar matter by the fast and energetic ejecta can
be also much brighter than that in the previous
study~\cite{HotokePiran,Hotoke18}. We also plan to quantitatively
explore this signal using our numerical models.

\vspace{3mm}
\acknowledgments

We thank Loren Held, Kenta Hotokezaka, Kyohei Kawaguchi, Kenta Kiuchi,
and Koh Takahashi for helpful discussions.  MS and SF thank Yukawa
Institute for Theoretical Physics, Kyoto University for their
hospitality during the first corona pandemic time in Germany, in which
this project was started.  This work was in part supported by
Grant-in-Aid for Scientific Research (Grant No.~JP20H00158) of
Japanese MEXT/JSPS.  Numerical computations were performed on Sakura
and Cobra clusters at Max Planck Computing and Data Facility.


\end{document}